\documentclass[%
 aps,
 prb,%
 amsmath,amssymb,
reprint,%
superscriptaddress,
]{revtex4-1}

\usepackage{graphicx}
\usepackage{dcolumn}
\usepackage{bm}
\usepackage[mathlines]{lineno}
\usepackage{amsmath,esint}
\usepackage{xcolor}
\usepackage{url}
\usepackage{pbox}
\usepackage[section]{placeins}

\usepackage{array}

\newcolumntype{M}[1]{>{\centering\arraybackslash}m{#1}}
\newcolumntype{N}{@{}m{0pt}@{}}

\usepackage[caption=false]{subfig}

\begin{document}


\title[Sample title]{Thermal runaway of metal nano-tips during intense electron emission}

\author{A. Kyritsakis}
\email{andreas.kyritsakis@helsinki.fi}
\affiliation{Deparment of Physics and Helsinki Institute of Physics, University of Helsinki, PO Box 43 (Pietari Kalmin katu 2), 00014 Helsinki, Finland}

\author{M. Veske}%
\affiliation{Deparment of Physics and Helsinki Institute of Physics, University of Helsinki, PO Box 43 (Pietari Kalmin katu 2), 00014 Helsinki, Finland}

\author{K. Eimre}
\affiliation{Intelligent Materials and Systems Lab, Institute of Technology, University of Tartu, Nooruse 1, 50411 Tartu, Estonia}

\author{V. Zadin}
\affiliation{Intelligent Materials and Systems Lab, Institute of Technology, University of Tartu, Nooruse 1, 50411 Tartu, Estonia}

\author{F. Djurabekova}
\affiliation{Deparment of Physics and Helsinki Institute of Physics, University of Helsinki, PO Box 43 (Pietari Kalmin katu 2), 00014 Helsinki, Finland}

\date{\today}

\begin{abstract}

When an electron emitting tip is subjected to very high electric fields, plasma forms even under ultra high vacuum conditions.
This phenomenon, known as vacuum arc, causes catastrophic surface modifications and constitutes a major limiting factor not only for modern electron sources, but also for many large-scale applications such as particle accelerators, fusion reactors etc. 
Although vacuum arcs have been studied thoroughly, the physical mechanisms that lead from intense electron emission to plasma ignition are still unclear.
In this article, we give insights to the atomic scale processes taking place in metal nanotips under intense field emission conditions. We use multi-scale atomistic simulations that concurrently include field-induced forces, electron emission with finite-size and space-charge effects, Nottingham and Joule heating. 
We find that when a sufficiently high electric field is applied to the tip, the emission-generated heat partially melts it and the field-induced force elongates and sharpens it.
This initiates a positive feedback thermal runaway process, which eventually causes evaporation of large fractions of the tip.
The reported mechanism can explain the origin of neutral atoms necessary to initiate plasma, a missing key process required to explain the ignition of a vacuum arc.
Our simulations provide a quantitative description of in the conditions leading to runaway, which shall be valuable for both field emission applications and vacuum arc studies.

\end{abstract}

\maketitle

\section{Introduction}

Field electron emission plays a crucial role in numerous modern technological applications, from electron microscopy to flat displays \cite{egorov2017field}. 
However, if the applied field at an emitting cathode is increased beyond a limit, violent discharges in the form of arcs appear in the vacuum.
Vacuum arcs, also known as vacuum breakdowns, play a significant role in various technological applications by being either exploitable or highly undesirable.
For example, they can be used in ion sources \cite{MEVVA1985} or in physical vapor deposition \cite{Anders}.
On the other hand, they cause catastrophic emitter failure in electron sources \cite{Anders_PRL88,Dyke1953Arc,Descoeudres2009_dc}.
Moreover, they hinder the function and limit the performance of various vacuum devices that require high electric fields, such as fusion reactors \cite{McCracken1980}, vacuum interrupters \cite{slade2007} and powerful linear accelerators to be used in particle colliders for new insightful experiments at CERN (CLIC) \citep{clic}. 

Over decades, vacuum arcing strongly attracted the attention of researchers from different fields \cite{Dyke1953Arc, Dyke1953I,DolanII,Anders_PRL88,Anders_PRL93,latham1995,Anders,Descoeudres2009}, since various physical phenomena are involved. 
However, the mechanisms of plasma onset are still under debate. 
During an arc, plasma is ignited in the vacuum and burns until the available energy from the power source is consumed.
It is well-known since the 1950's \citep{Dyke1953I,DolanII,Descoeudres2009,Anders} that vacuum arcs appear after intense field electron emission, but the physical mechanism that leads from the latter to plasma ignition is not yet fully understood. 

Recent experimental studies show that the improvement of surface and vacuum quality cannot diminish the probability of vacuum arcing \cite{Degiovani_conditioning}.
Moreover, arcs were found to appear on single metal tip cathodes when their current density exceeded a critical value even for well-controlled emission conditions \cite{Dyke1953Arc}. These indicate the existence of an inherent mechanism of surface response that initiates breakdown.
One hypothesis commonly used to explain the plasma build-up is the "explosive emission"\citep{Anders, Mesyats_Ecton, mesyats1993ectons, Mesyats2005}, which assumes that an instant explosion of the emitting tip leads to plasma formation.
However, this hypothesis is mainly based on phenomenological considerations, since the atomic level insight in such a complex phenomenon as vacuum arcing was not available. 

Here we present multi-physics atomistic simulations that reveal a thermal runaway process on emitting nano-tips. The latter is based on the gradual deformation of an emitting tip, due to the field-induced forces. 
The simulations show that both high current density and sufficiently large tip size are needed to initiate the melting at its apex. After that, the field-induced forces gradually deform the tip, elongating and sharpening its apex in a process similar to the Taylor cone formation in liquid metal ion sources \cite{krohn1975ion, Wagner1982hydrodynamics, Swanson_LMIS}. 
This process, in combination with the decrease of the electrical and thermal conductivities at high temperatures, leads to a positive feedback thermal runaway mechanism.
Eventually, large fractions of the emitter evaporate in the form of neutral atoms, but also as charged nano-clusters. 
The total number of evaporated atoms (both isolated and clustered) is compatible with the minimum evaporation rate required to ignite plasma, as found by recent Particle-In-Cell (PIC) calculations \cite{ArcPIC_1d, arcPIC}.

\section{Method} \label{sec:method}

In order to study the atomic level processes leading to the initiation of plasma, we perform  Molecular Dynamics (MD) simulations. 
However, the thermal runaway process involves various physical phenomena, which classical MD cannot describe implicitly due to its atomistic nature.
The electric field interacts with the material, inducing charges and forces on the surface atoms and generating electron emission. The latter heats the nano-tip due to the Nottingham \cite{Nottingham,NottingCharb} and Joule effects.
The combination of the above causes significant changes in the structure and  geometry of the material and hence they have to be quantified and included in the calculations. 

\begin{figure}[htbp]
	\centering
    \includegraphics[width=.99\linewidth]{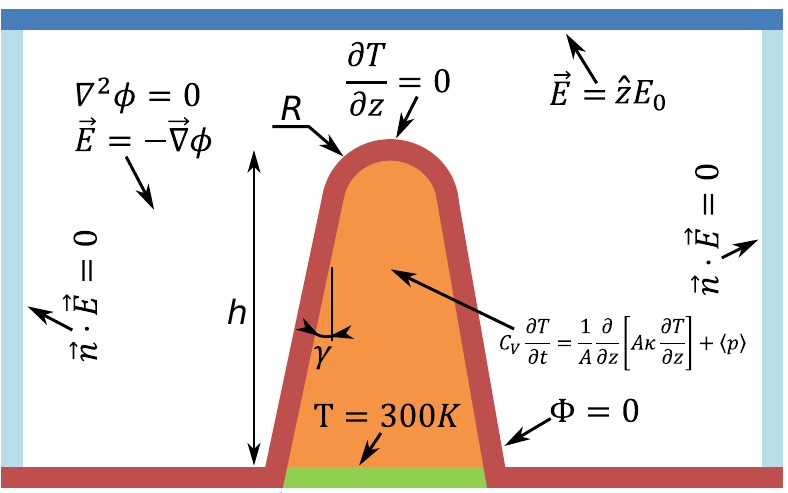}
    \caption{Schematic of the partial differential equations and their corresponding boundary conditions.}
    \label{fig:schematic}
\end{figure}

In this article we propose a multi-physics model, which extends the capability of the classical MD method by solving the electrostatic, electron emission and heat diffusion equations concurrently for a dynamically evolving nano-emitter shape. 
In the next three sections we explain how these effects are incorporated in the MD simulations.
In appendix \ref{sec:flowchart} a flowchart of the whole simulation algorithm is given. 
Figure \ref{fig:schematic} illustrates a schematic representation of the tip geometry that along with a compilation of the partial differential equations with their corresponding boundary conditions used to obtain the electric field and temperature distributions.


\subsection{Electric field} \label{ssec:field}

In order to introduce the electrostatic phenomena, we have updated our HELMOD model \cite{Djurabekova2011} that couples MD simulations with electric field calculations, and includes the charge-induced forces in the MD interactions.
Although HELMOD provided many interesting insights into the interaction of high electric fields with metal surfaces, the rigid grid used to calculate the electric field imposed critical limitations for large systems, dynamically evolving at high temperatures. 
To overcome this limitation we used our recently developed model FEMOCS \cite{FemocsIVNC,VekseDynamic_arxiv} to calculate the electric field distribution around the tip on a flexible mesh and integrate this information into HELMOD.

As HELMOD, FEMOCS also solves the Laplace equation 
\begin{equation} \label{eq:Laplace}
	\nabla^2 \Phi = 0 \textrm{,}
\end{equation}
but uses the Finite Element Method (FEM) on an unstructured mesh. 
This mesh is generated automatically from the atomic positions in order to follow the varying nano-tip shape. A Dirichlet boundary condition
\begin{equation} \label{eq:Dirichlet}
	\Phi=0 
\end{equation}
is then applied on the nano-tip surface and a Neumann
\begin{equation} \label{eq:Neumann}
	\nabla \Phi = \hat{z}E_0
\end{equation}
at the top of the simulation box, where $E_0$ is the applied macroscopic field.
Once the electrostatic potential $\Phi$ and field 
\begin{equation} \label{eq:Nabla}
	\vec{E} = - \nabla \Phi
\end{equation}
are obtained, we calculate the charges $q_i$ induced in each atom due to the field (details on how this charge is calculated are given in appendix \ref{sec:charges}).
Given the charge $q_i$ on each atom, we calculate and add to the MD interaction the Lorentz force
\begin{equation} \label{eq:Forces}
	\vec{F}_i = \frac{1}{2} q_i \vec{E}_i \textrm{.}
\end{equation}
The $1/2$ term stands for the fact that the atoms are exposed to the electric field only from the one side of the material-vacuum surface.
The inter-atomic Coulomb forces are also calculated and included in the MD interaction by the method described in ref. \cite{Djurabekova2011}.

The electric field effects are calculated concurrently with MD, so that the field distribution evolves according to the emitter shape. When the emitter shape changes significantly, the electric field distribution is recalculated.
At every timestep, the position of the atoms is compared to the last full-calculation step. 
If the root-mean-square (RMS) average of the atomic displacement (compared to the last field calculation step) is smaller than 0.38$\textrm{\AA}$, the last calculated field solution is reused.
This technique is used in order to keep the computational time of our simulation feasible. 
Choosing the value for the threshold is a trade-off between computational efficiency and approximation accuracy.
The aforementioned threshold was used because it reduced the CPU time by about two orders of magnitude, while the level of accuracy of the field calculations deteriorated only by 2\% as compared to the accuracy o
A full analysis of the CPU time and field calculation error as a function of the threshold can be found in ref. \cite{VekseDynamic_arxiv}.

\subsection{Electron emission} \label{ssec:emission}

Our previous calculations \cite{Parviainen2011, Eimre2015} showed that the thermal effects caused by electron emission play an important role in the nanotip evolution under high fields.
The classical electron emission equations \cite{Schottky1923, FN1928, MurphyG} are inadequate to describe combined thermal-field emission \cite{Jensen2006} from nanometrically sharp emitters \cite{CutlerAPL, KXnonfn,KXGTF}. However, our recently developed computational tool GETELEC \cite{GETELECpaper} provides with the means to consistently and efficiently calculate the electron emission current and Nottingham heat distributions from sharp nano-tips, even at temperatures beyond the melting point.
We also note that the local fields and the current densities calculated here are close to the Space Charge (SC) limit; thus the SC effect \cite{Child_SC} is also included in our model. 

Calculating fully the charge density distribution induced by the emitted electrons in three dimensions (3D) is a complicated and computationally expensive calculation \cite{Jensen1999}, which is usually done with the PIC method \cite{chen2009space, Uimanov2011}.
Nevertheless, Forbes \cite{ForbesSpace} claimed that the standard 1D SC model for field emission \citep{BarbourSC} can be used to obtain the reduction of the surface field if a correction factor $\omega$ is introduced to account for the non-planar nature of the emitter. 

According to the standard 1D SC model, the local field on the emitter (uniform in 1D) is found as $F = \theta F_L$, where $\theta$ is the dimensionless field-lowering factor and $F_L$ is the field found by the Laplace equation, i.e by ignoring the SC effects.
$\theta$ is a function of the emitted current density $J$, the local field $F$ and the total applied voltage $V$, which is here an external parameter.

However, for a 3D nanometric emitter, $J$ and $F$ are not well-defined.
In a quasi-planar emitter, the values at the apex can be used, since the SC is localized around it in a region much smaller than the radius of the emitter, where these quantities are practically constant.
On the contrary, in a nanometric emitter the whole tip surface contributes to the SC, which means that the apex values are no longer representative. 

Here we will define the following representative values $F_r,J_r$  for the field and the current density correspondingly
\begin{eqnarray} \label{eq:repr}
\displaystyle
	J_r = \frac{\int_{S_h} ~ JdS}{ \int_{S_h} ~ dS} \nonumber \\
	F_r =  \frac{\int_{S_h} ~ FJdS}{ \int_{S_h} ~ JdS} \textrm{.} 
\end{eqnarray}
In eq. \eqref{eq:repr} the surface integrals are performed on the full-width-half-maximum emission surface $S_h$, i.e the surface where $J>\frac{1}{2}J_{max}$.
On this surface, $J_r$ is the mean value of $J$ and $F_r$ is the weighted mean value of $F$, with the weight being the emitted current density. 

The field suppression factor can then be obtained by solving the equation \cite{ForbesSpace}
\begin{equation} \label{eq:Barbour}
	9\zeta \theta^2 -3 \theta -4 \zeta +3 = 0 \textrm{,}
\end{equation}
with 
\begin{equation} \label{eq:zeta}
	\zeta \equiv \frac{1}{\epsilon_0} \sqrt{\frac{m}{2e}} \frac{J_r \sqrt{V}}{F_r^2} \textrm{,}
\end{equation}
where $m$ is the electron mass, $e$ the elementary charge and $\epsilon_0$ the vacuum permittivity.
 
The whole potential and field distributions on the emitter are then multiplied by $\theta$ and the current density distribution along with $J_r,F_r$ are recalculated according to the new  field distribution.
This procedure is repeated iteratively until self-consistency is reached, i.e. the $\theta$- suppressed field and the corresponding current density distributions give $J_r,F_r$ that satisfy eq. \eqref{eq:Barbour}, \eqref{eq:zeta}. 
Note that the obtained self-consistent $\theta$ is used to multiply the whole electric field distribution, thus affecting not only the electron emission, but the charge and force calculations described in section \ref{ssec:field}.
 
In order to validate our 1D space charge model, we compare its results to full PIC simulations performed by Uimanov \cite{Uimanov2011}.
We reproduced the solution of the Laplace equation for the geometries presented in ref. \cite{Uimanov2011}, and then used our model to obtain the field-lowering factor $\theta$, using a corrected applied voltage $\omega V$. $\theta$ is obtained for various values of the applied voltage and for two of the geometries simulated by Uimanov: "FE4" and "FE5". "FE4" is an emitter with 10nm radius and "FE5" with 1nm, thus being the most relevant to our simulated tips that have a radius $R=3$nm (see section \ref{ssec:setup}). 

\begin{figure}[htbp]
	\centering
    \includegraphics[width=\linewidth]{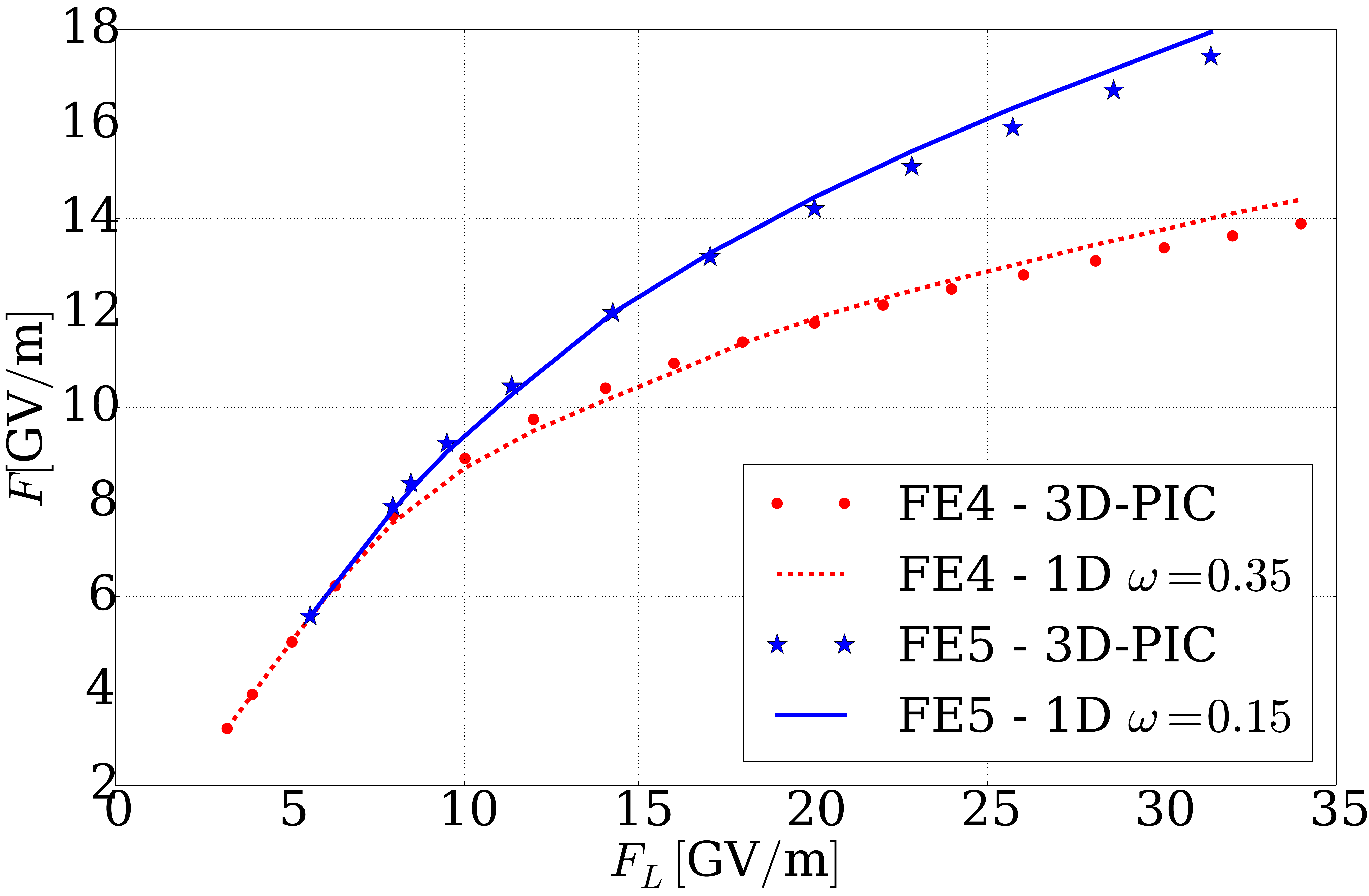}
    \caption{Local apex field including the SC $F$ versus the one ignoring SC $F_L$ for the two different emitter geometries simulated by Uimanov \cite{Uimanov2011}. Markers correspond to the results of ref. \cite{Uimanov2011} and the lines to the results obtained by our 1D SC model.}    \label{fig:SCcomp}
\end{figure}

In figure \ref{fig:SCcomp} we see the $F-F_L$ (values at the apex) curves reported in ref. \cite{Uimanov2011} (markers) along with the results of our model (lines) for the same geometries.
The corresponding fitted values for the correction factor $\omega$ are shown in the legends.
We see that our model is in very good agreement with the full 3D PIC calculation with an error of less than 1\%, at a range of $F_L$ up to about 35GV/m. 

In view of the above, the 1D SC model described above is a good approximation for nanometric size emitters, given that a valid correction factor $\omega$ is used.
Although we cannot obtain here its value for the specific geometry we simulate, the fitted values obtained above for similar hemisphere-on-a-cone geometries indicate that the used value $\omega = 0.25$ is reasonable for our nano-tip.
This value was chosen for $R=3$nm, by linearly interpolating between $\omega=0.15$ and $\omega = 0.35$, with respect to the corresponding $\log(R)$.

\subsection{Heating} \label{ssec:heat}

In order to obtain the temperature distribution $T(x,y,z;t)$ on the emitter, we solve the heat diffusion equation \cite{bejan2003heat}
\begin{equation}
	\label{eq:3Dheat}
	C_V \frac{\partial T}{\partial t} = \nabla \cdot \left(\kappa(T) \nabla T \right)+ p \textrm{,}
\end{equation}
where $C_V$ is the volumetric heat capacity, $\kappa(T)$ is the heat conductivity and $p$ is the local deposited volumetric heating power density (in $\textrm{W}/\textrm{m}^3$). 

The geometrical structures which we simulate, have one dominant dimension, i.e. their height $h$ is much longer than the lateral dimensions.
Hence eq. \eqref{eq:3Dheat} can be reduced to one dimension
 \begin{equation}
        \label{eq:1Dheat}
        C_V \frac{\partial T}{\partial t} = \frac{1}{A(z)} \frac{\partial}{\partial z} \left[ A(z) \kappa(z) \frac{\partial T}{\partial z} \right]  + \langle p \rangle \textrm{,}
 \end{equation}
where $A(z)$ is the cross-sectional area of the tip at a given height $z$ and $\langle p \rangle$ is the mean deposited heat on the corresponding cross-section. 
Details on the exact definition of $p, \langle p \rangle$ and the derivation of \eqref{eq:1Dheat} can be found in appendix \ref{sec:1dheat}.

The above equation is solved using the Finite Difference Method (FDM) with an explicit Euler scheme.
We divide the tip in $N$ finite slices of width $\Delta z$, that coincide with the slices of the rectangular grid used for the charge calculation (see appendix \ref{sec:charges}).
On each surface grid cell, the current density and the deposited Nottingham heat surface density are calculated as described in section \ref{ssec:emission}.
Then the total current and total Nottingham heating power contributions of each cell ($i_c$ and $p_{N_c}$ correspondingly) are calculated by multiplying the densities with the cell face areas, similarly to the charge calculation described by equations \eqref{eq:Gauss} and \eqref{eq:Qatom}.
We consider that the emission quantities have the same direction as the electric field of the cell.

The total current flowing through the cross-section of the $k$-th ($k=0$ at the apex) slice is cumulative from the apex to the base as dictated by the continuity equation, i.e
\begin{equation}
	I_k = I_{k-1} + \sum_{c \in S_k}~i_c
\end{equation}
where $S_k$ denotes the cells of the $k$-th slice that are exposed to vacuum.
The total heat deposited on the $k$-th slice is the sum of the Joule and the Nottingham heat components, i.e 
\begin{equation}
	\label{eq:power}
	P_k = \frac{{I_k}^2}{A_k \sigma_k} \Delta z + \sum_{c \in S_k}~p_{N_c} \textrm{.}
\end{equation}

For the Joule heating, we need to obtain the electric conductivity $\sigma$ which depends on the local temperature and the size of the nano-tip.
For a given temperature $T$, we obtain $\sigma(T)$ by linearly interpolating tabulated values found in the literature \cite{Matula1979,Gathers1983} for the resistivity $1/ \sigma$.

$\sigma(T)$ is capped at the lowest available value of 3500K, i.e. $\sigma(T>3500\textrm{K})=\sigma(3500\textrm{K})$. 
Furthermore, the mean free path of the electrons in the material is decreased due to the nanometric size of the nano-tip.
To correct our $\sigma$ values for this finite-size effect, we use the simulation method and tool of Yarimbiyik et. al. \cite{Yarimbiyik2005}.
The latter can calculate the mean free path reduction for a given nano-wire diameter.
The value we use for the latter is the mean diameter along the tip for the initial geometry. 

The heat conductivity $\kappa$ is calculated from $\sigma$ according to the Wiedemann -- Franz law $\kappa = LT\sigma$. The Lorentz number $L$ is known to be reduced for nanometric size structures \cite{Nath1974}.
We used its smallest value $L = 2.0 \times 10^{-8} \textrm{W} \Omega \textrm{K}^{-2}$ reported in ref. \cite{Nath1974} for a Cu film of $40 \textrm{nm}$ thickness. We note that for a nano-wire of smaller size such as the ones we simulate, $L$ might decrease even more, which will promote the thermal runaway even for smaller fields or emitter sizes.

The Euler scheme for equation \eqref{eq:1Dheat} is $T_k(t+\delta t) = T_k(t) + \Delta T_k$. Discretizing equation \eqref{eq:1Dheat} yields
\begin{equation}
	\label{eq:FDM}
	\begin{split}
	\Delta T_k = \frac{\delta t}{C_V A_k \Delta z} \left[ \frac{\kappa_k A_k}{\Delta z} \left( T_{k+1}+T_{k-1} - 2T_k \right) + \right. \\
	 \left. \frac{1}{4 \Delta z} \left( A_{k+1} \kappa_{k+1} - A_{k-1} \kappa_{k-1} \right) \left( T_{k+1} - T_{k-1} \right) + P_k \right]
	\end{split}
\end{equation}
where the subscript denotes the slice number for all quantities. 
A constant Dirichlet boundary condition is applied at the bottom of the tip, i.e $T_N = 300$K. Note that in \eqref{eq:FDM} values for $k=-1$ are needed to obtain $\Delta T_0$. Thus we assume a virtual point for which $A_{-1} = 0$ and $T_{-1} = T_0$. The latter corresponds to a Neumann boundary condition of zero heat flux at the apex.

The forward difference Euler scheme of eq. \eqref{eq:FDM} is numerically unstable if $\delta t$ and $\Delta z$ are not sufficiently small. Here we used $\delta t = 0.06 \textrm{fs}$ and $\Delta z \approx 1.8 \textrm{\AA}$. 
$\Delta z$ is determined by the rectangular grid which might change as the MD simulation evolves and is approximately half a lattice constant. 
Our tests showed that these values are small enough to safely avoid numerical instabilities.
Besides, the computational cost of iteratively updating eq. \eqref{eq:FDM} is negligible comparing to the rest of the simulation, thus there is no need to optimize their values as long as they are small enough.

The conductivities $\sigma_k, \kappa_k$ are updated at each heating time step $\delta t$ according to the current $T_k$. On the other hand, $A_k$ and $P_k$ are recalculated and updated when either the field distribution is recalculated (meaning that the atoms have been displaced more than a limit) or $T(z)$ has an RMS difference greater than 1\% compared to the last step when $A_k$ and $P_k$ where obtained. 

Finally, the resulting temperature distribution $T(z)$ is used to control the temperature of the MD system.
The velocities of the atoms residing inside the $k-$th slice are scaled according to a Berendsen control scheme \cite{Berendsen} with control temperature $T_k$ and relaxation time $\tau = 1.5\textrm{ps}$.
This value is much smaller than the relaxation time of the heat equation, but also big enough to avoid artefacts appearing in MD when intense velocity scaling is applied. 

The above 1D model has the advantage of its simplicity and computational efficiency.
In order to validate the model, we compare its results with our previous 3D Finite Element Method (FEM) model \cite{Eimre2015} (see appendix \ref{sec:1dvalid} for details). 
This comparison shows that the 1D heat model described above produces results that are in excellent agreement with ones from 3D FEM.

\subsection{MD simulation set-up} \label{ssec:setup}
\begin{figure}[htbp]
	\centering
    \includegraphics[width=.99\linewidth]{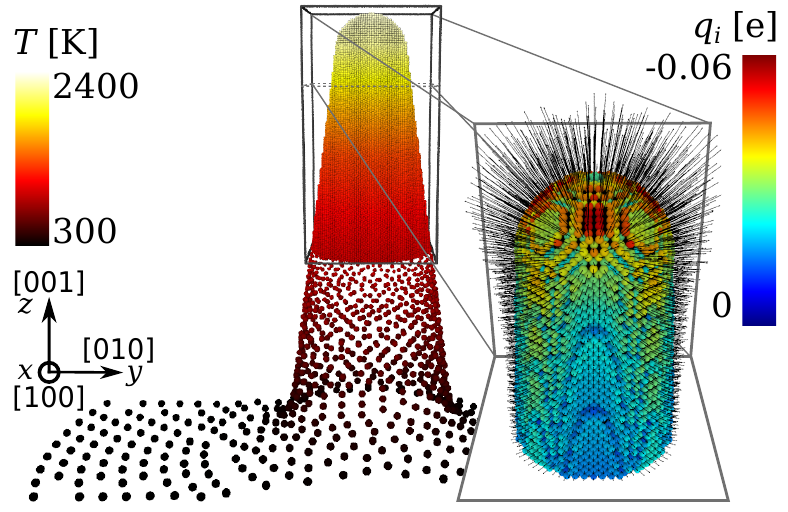}
    \caption{Simulated tip shape illustrated by the MD atoms (in the box) and the constant "virtual" atoms (out of the box). The color coding represents the steady-state temperature distribution for constant shape. The coordinate reference system is placed in the middle of the tip at the bottom of the MD domain. The inset illustrates the atom charges $q_i$ by the color coding and, the corresponding forces $\vec{F}_i$ by the black-line arrows.}
    \label{fig:resheat}
\end{figure}

We simulate a conical Cu nanotip whose major axis aligned with the $\langle 100 \rangle$ crystallographic direction (see figure \ref{fig:resheat}).
The cone is terminated with a hemispherical cap of a radius $R = 3nm$.
The tip has a total height $h = 93.1\textrm{nm}$ and a full aperture angle of $3^o$.
In order to keep the computational time feasible, only the upper half of the tip is fully simulated with MD.
According to our estimations, only this part is significantly heated, providing high kinetic energies to the atoms.
The bottom half maintains practically a constant shape and we consider it fixed; the field and temperature calculations are extended to the continuous limit in that region.
Thus the total number of simulated atoms is 206000 and the MD simulation box has a size of $11.64 \times 11.64  \times 46.6$nm (shown with grey lines in figure \ref{fig:resheat}).

However, the electrostatic and heat equations are solved on the whole tip domain, which includes an extension of the MD system shown by the "virtual" particles lying out of the box in figure \ref{fig:resheat}. These points are used to generate the FEM mesh on a $280 \times 280  \times 560$nm box.
This size is sufficient to assume periodic boundary conditions at the sides and a Neumann boundary condition at the top. 
Only a small part of this box is shown in the figure. A macroscopic electric field $E_{0} = 0.8\textrm{GV/m}$ with a total voltage $V_{0} = 3$kV is applied instantly after the MD system is run to relax for 0.5ps. These values for $E_0,V_0$ are relevant to the surface fields used in modern high-gradient accelerating structures \cite{Degiovani_conditioning,Wu_high-gradient}. 

For the MD simulations we used two different inter-atomic potentials with a constant timestep $\Delta t = 4.05$fs. The Sabochick and Lam (SL) EAM potential \cite{SL_EAM_CU} and the one from Mishin et. al. \cite{Mishin}. In order to take into account the stochastic nature of the process, we repeated the same simulation 7 times for each potential. 
Each repetition was initialized with different atomistic velocities (randomly sampled from the Maxwellian distribution with $T=300\textrm{K}$) by giving different seeds to the corresponding random number generator.

Since our tip is isolated and continued to the bottom, no periodic boundaries were applied. The two bottom atomic layers of the MD systems were fixed to obtain a smooth connection to the rest of the tip and avoid translational motion of the whole structure. Finally, the ten bottom layers above the two fixed were controlled to linearly increasing temperatures from 0 to $T(z)$ in order to avoid possible artefacts caused by extreme velocity mismatch. The temperature on the remaining of the tip was controlled according to the calculated $T(z)$.

We note that the total computational time needed for the simulation is quite high. Every simulation runs on a single core and it needed approximately 10 days to run the presented 350ps simulation time either on a standard workstation or cluster. The memory consumption of the simulation does not exceed 2GB.  

\section{Results} \label{sec:results}

After the electric field is applied, the maximum local field at the apex as calculated by the Laplace equation is $E_{max} = 17.5 \pm 0.06 \textrm{GV/m}$.
However, such a high field induces an emitted current density with an average value at the bottom of the MD simulation domain (middle of the conical tip) $J_{avg} = 3.51 \pm 0.02 \times 10^{-5} \textrm{Anm}^{-2}$.
This will suppress the fields due to the SC effect by a factor of $\theta = 0.69 \pm 0.001$ and converge to a final current density of $J_{avg} = 4.5 \pm 0.03 \times 10^{-6} \textrm{Anm}^{-2}$. The inset of figure \ref{fig:resheat} shows a closer view on the tip apex with the color coding corresponding to the calculated field-induced charges $q_i$ for each atom and the black arrows corresponding to the forces $\vec{F}_i$. 
All the numerical results reported throughout the next two sections in the format "$x \pm \delta x$" or with plot error bars correspond to the mean values and standard errors as obtained from the various simulation runs with different random seeds.

\begin{figure}[htbp]
	\centering
	\includegraphics[width=.99\linewidth]{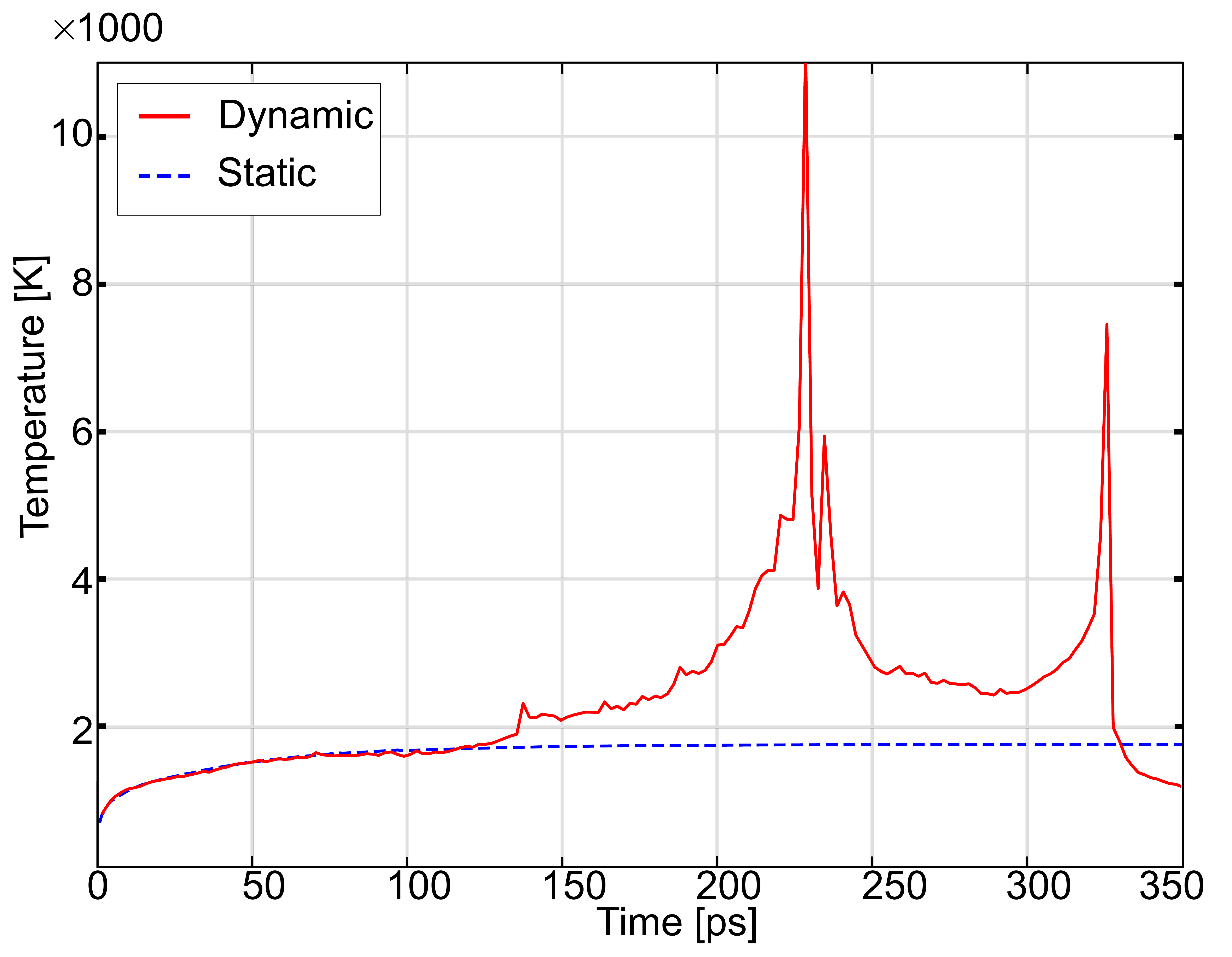}
    \caption{Time evolution of the maximum temperature on the tip of a constant shape (dashed blue line) and of a variable shape as calculated by the multi-scale atomistic method (solid red line).} 
    \label{fig:temperatures}
\end{figure}

Figure \ref{fig:temperatures} demonstrates the evolution of the maximum temperature $T_{max}$ on the nano-tip for one of the simulations with the SL potential.
In the initial geometrical configuration, the deposited heat power density is very high (reaches $3 \pm 0.1 \times 10^{-5} \textrm{Wnm}^{-3}$ at the apex) and is dominated by the Nottingham effect.
If the heat equation is solved under constant geometry, the heat is rapidly dissipated towards the bulk due to the high thermal conductivity of Cu and the temperature distribution converges to a steady state, with a maximum value of about $1820 \pm 24$ K at the apex.

\begin{figure}[htbp]
	\centering
    \subfloat[]{\label{fig:height} \includegraphics[width=.99\linewidth]{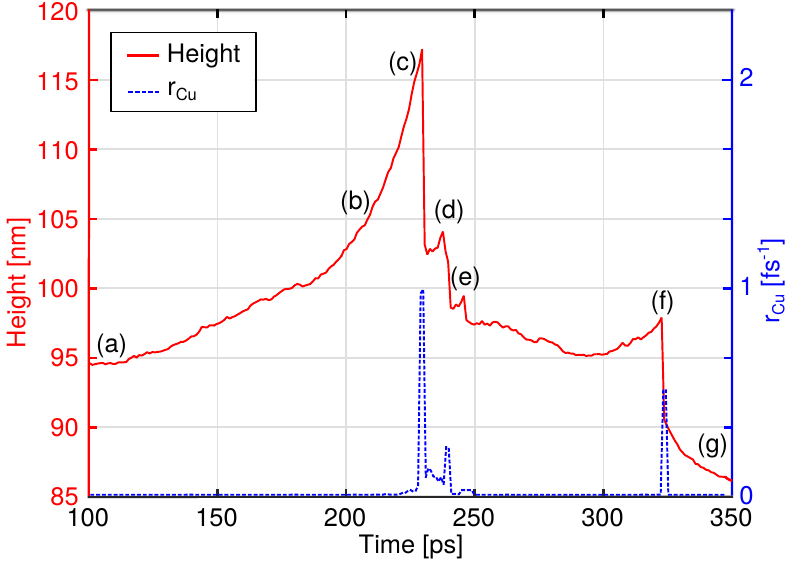}}

	\subfloat[]{\label{fig:frames} \includegraphics[width=.99\linewidth]{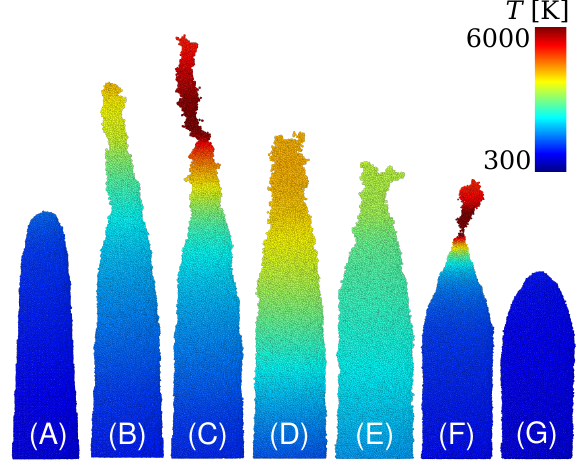}}

	\caption{\label{fig:evolution} (a) Evolution of the height of the tip (red line, left axis) and the Cu atoms evaporation rate (blue dashed line, right axis). The designated points (a-g) correspond to the frames depicted in figure (b). The color coding of (b) corresponds to the local temperature. All evaporated atoms and nano-clusters have been removed. The full animation is available in the supplementary material.}
\end{figure}

On the other hand, if we let the tip shape evolve according to the multi-scale method described in section \ref{sec:method}, we obtain a completely different picture (solid red line). The flexible shape of the molten material on the top of the tip significantly affects the field and temperature distributions; this causes a thermal runaway process leading to extremely high temperatures.

\begin{figure}[h]
	\centering
    \subfloat[]{\label{fig:Heat} \includegraphics[width=.99\linewidth]{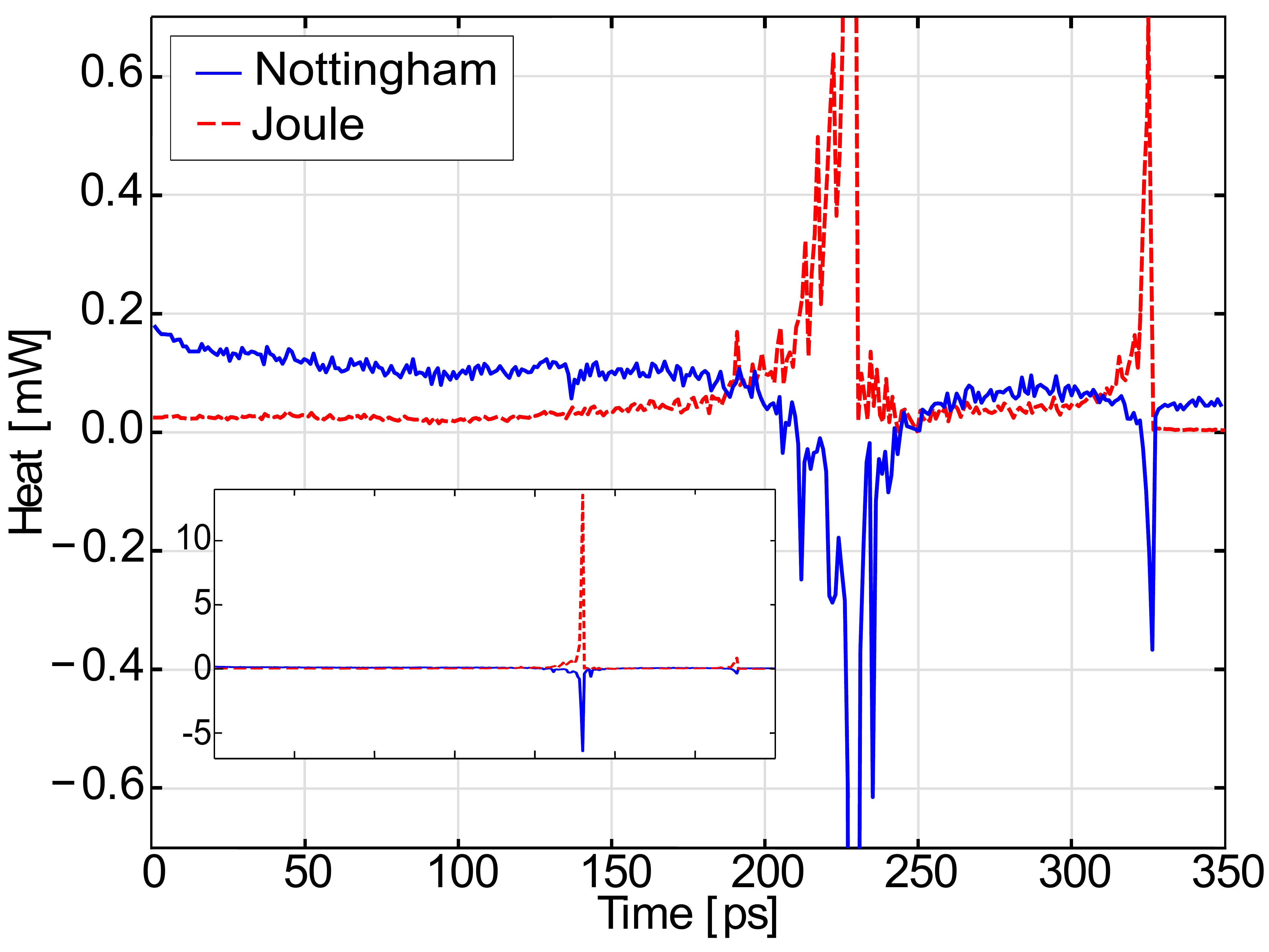}}

	\subfloat[]{\label{fig:Field} \includegraphics[width=.99\linewidth]{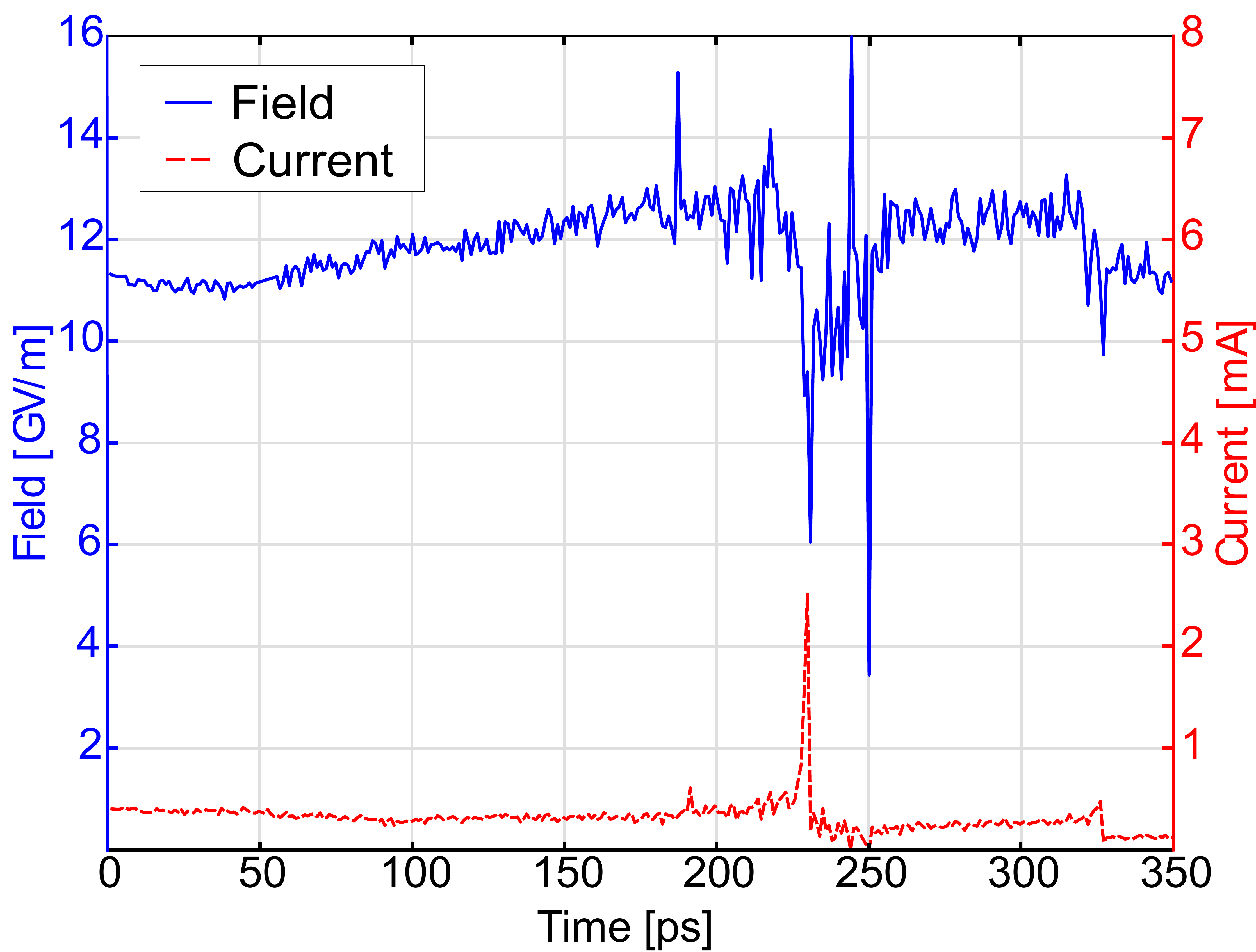}}

	\caption{\label{fig:hfc} Time evolution of: (a) the total deposited Nottingham heating power (solid blue line), Joule heating power (dashed red line) and (b) the representative electric field $F_r$ (solid blue line, left axis) and the total emitted current (dashed red line, right axis). The inset of (a) shows the same data in their full range.}
\end{figure}

This shape-related runaway process is illustrated in figures \ref{fig:evolution} and \ref{fig:hfc} for the same simulation. The solid red line of figure \ref{fig:height} shows the time evolution of the height of the tip. Figure \ref{fig:frames} demonstrates the snapshots (a-g) of the tip evolution at the time steps designated on the graph respectively. Figure \ref{fig:Heat} shows the corresponding evolution of the deposited heat components (dashed red for Joule and solid blue for Nottingham) and figure \ref{fig:Field} gives the evolution of the representative field $F_r$ of eq. \eqref{eq:repr} (solid blue line) and the total emitted current (dashed red line - right axis) for the same simulation.

The tip shape initially stays approximately constant ($t<100 \textrm{ps}$), while its temperature gradually increases towards its steady-state value of $T_{max} = $1820K. 
In this stage the Nottingham heating power is dominant (about an order of magnitude higher than Joule), but as the temperature increases it slowly reduces.
The steady-state temperature exceeds the melting point at the apex region and the corresponding atoms become mobile. 
Then the field-induced forces can pull them upwards, thus sharpening and elongating the tip.
This leads to a slow increase of the local fields, as shown in the figure \ref{fig:Field} ($t > 100$ps). 
The increased field induces higher forces that elongate and sharpen the tip even further. 

At this stage the total emitted current stays roughly constant (see figure \ref{fig:Field}) because the emission is deeply in the space-charge limited regime.
Nevertheless, the increase of the tip hight in combination with the increased resistivity hinders the heat conduction and causes an increase in the Joule heat and temperature.
After about $t=150$ps, the above processes have formed a positive feedback loop leading to thermal runaway.
The system height, field, temperature and Joule heat components constantly increase feeding the increase of each other. 

Shortly after $t=200$ps the deformed tip has exceeded 3000K (frame b in figure \ref{fig:evolution}).
At this temperature the Nottingham effect is inverted, producing cooling instead of heating due to the increased thermionic component of the emission (blue curve in figure \ref{fig:Heat} becomes negative).
However, the high temperature also results in significantly increased resistivity.
Furthermore, the field-induced forces cause neck-thinning which becomes more pronounced as the simulation evolves (frame c).
This thinning confines the total emitted current in a small cross-sectional area, thus producing a high local current density. 
The combination of the above produces high Joule heating, which exceeds the Nottingham cooling and leads to a rapid increase of the local temperature exceeding 10000K for a very short time. 
The emitted current at this point increases abruptly because of the thermionic component of the emission.
We see that the corresponding field decreases rapidly at the same point due to the space charge limitation, but this does not affect the emission due to its thermionic nature at that point.

Eventually, the high temperature in combination with the field-induced forces cause detachment of the upper part of the tip, creating an evaporated charged nano-cluster.
The latter is charged due to the partially charged atoms on its surface.
In the one presented in figure \ref{fig:frames} it consists of 1784 particles that bear a total charge of -23.8e. 

Although the full behaviour of evaporated nano-clusters requires further investigation (interaction with the field, collision with the evaporated atoms, emitted electrons, etc.), for the purpose of this work we consider them instantly removed and continue the simulation with the remaining tip. This is justified by the fact that they will be rapidly accelerated by the field towards the anode due to their negative charge.

All evaporated atoms and clusters are marked and removed by a cluster analysis algorithm \cite{ester1996density} implemented in FEMOCS \cite{VekseDynamic_arxiv}, with a distance cut-off equal to the one of the MD potential ($4.94 \textrm{\AA}$ for SL and $5.5 \textrm{\AA}$ for Mishin).
Figure \ref{fig:frames} shows only the non-evaporated atoms.
Note that we have not explicitly forced any evaporation event.
They appear naturally in the MD simulation due to the high temperature and the field-induced forces.

The dashed blue line of figure \ref{fig:height} (right axis) demonstrates the evolution of the evaporation rate of Cu atoms. The peaks in that line correspond to the evaporation of nano-clusters.
Shortly after the evaporation of the first cluster, the tip is very hot and many atoms evaporate either as isolated or clustered. The peaks on frames (d) and (e) in the height and evaporation rate graphs correspond to two relatively large clusters being detached.

After $t = 250$ps the tip height and temperature gradually decrease, and the system returns to a state similar to the one before $t=200$ps. 
However, at about $t = 290$ps the runaway process is reinitialized in the remaining tip until a new large cluster is evaporated at frame (f).
After that, the process finally stops as the tip blunts and cools down (g).
The full animation of the evolution of the tip can be found in the supplementary material.

Although the runaway and the evaporation appeared at different times and the tips took various shapes in the 14 simulation runs with different random seeds, the steps described above remained qualitatively similar. 
After a certain time a runaway process leads to a large cluster evaporation (corresponding to frame (c) of figure \ref{fig:evolution}) which is followed by smaller runaway--evaporation events (frames (d), (e) and (f)). 

\begin{figure}[h]
	\centering
    \subfloat[]{\label{fig:events} \includegraphics[width=.99\linewidth]{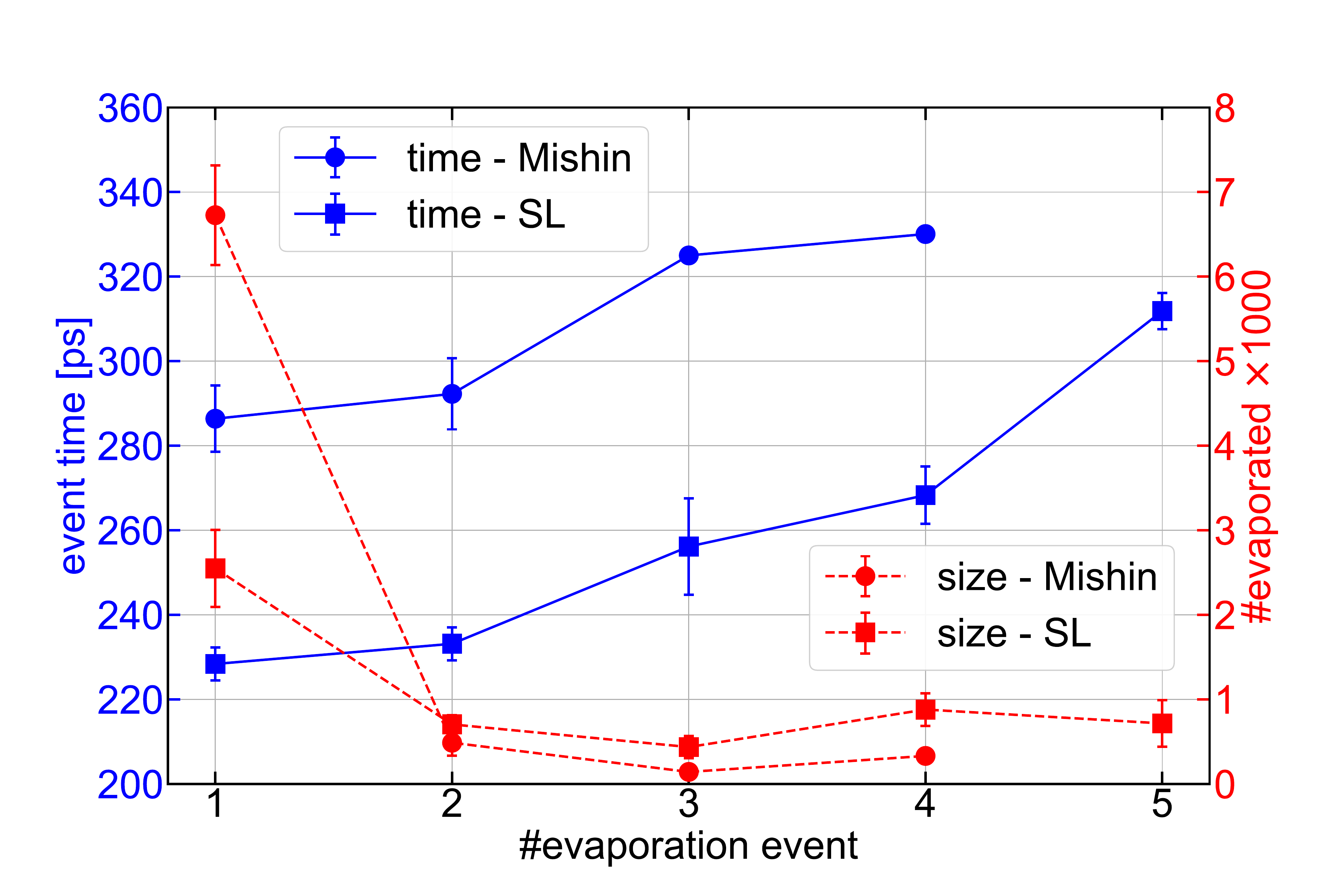}}

	\subfloat[]{\label{fig:evaps} \includegraphics[width=.99\linewidth]{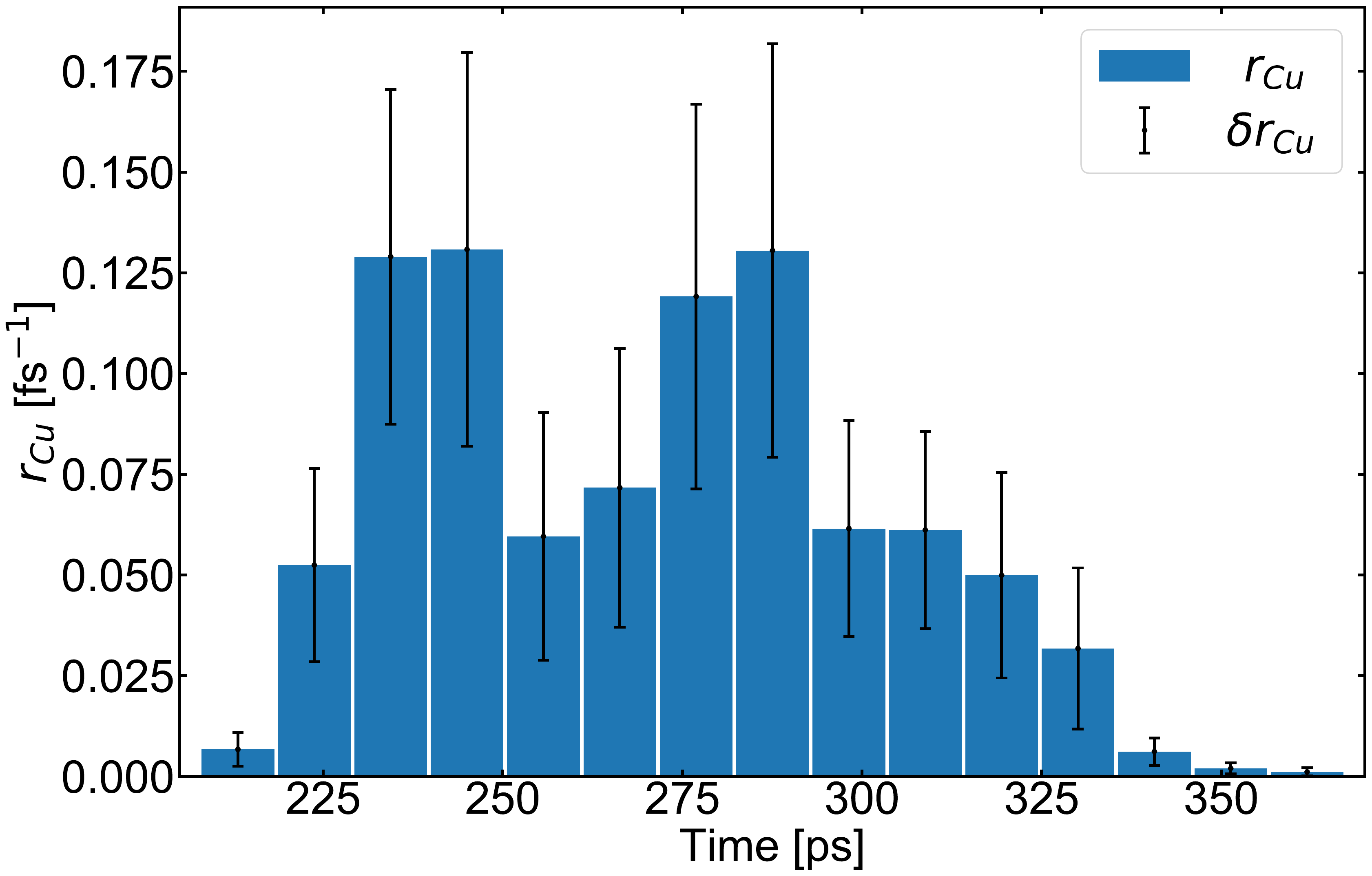}}

	\caption{\label{fig:stats} (a) Time (left axis - blue solid lines) that each evaporation event appears, along with the number of atoms of the corresponding evaporated nanocluster (right axis - red dashed lines). Round markers correspond to calculations with the Mishin potential square ones to the SL. Events \#3 and \#4 appeared only for one of the runs with the Mishin potential, hence the absence of error bars. (b) Time--averaged evaporation rate (bars). The width of the bar corresponds to the time interval over which the evaporation has been averaged. The error bars correspond to the standard error of all the runs for both potentials.}
\end{figure}

The statistical behaviour of the events is demonstrated in figure \ref{fig:events}, where we plot the time when each event appeared along with the size of the corresponding evaporated nanocluster. 
The behaviour of these events was slightly different for the two potentials, hence we plotted their statistics separately. 
The SL potential gives the first runaway significantly earlier than the Mishin one, while the size of the corresponding nano-cluster is smaller. 
However, after the first event, the SL simulations produce more and larger events, because the corresponding height has not been reduced as much as for the Mishin.
Thus, the mean number of events is $4 \pm 0.3$ for the SL potential and $2.1 \pm 0.3$ for the the Mishin (there is no simulation with 5 events for Mishin).

We see that the events shown in figure \ref{fig:events} are quite localized in time (cf. small corresponding error bars), which implies a relatively small stochasticity of the runaway-evaporation process.
This means that the latter substantially differs from the usual thermal evaporation, that would appear as purely random.
Thus, although the runaway is a thermally-activated process, it is mainly driven by the electric field.

The events described above, along with the atomic thermal evaporation create a statistical distribution for the evaporation rate, which is shown in figure \ref{fig:evaps}. Due to the sharp peaks of the various events, the mean rate is very noisy. For this reason we plot its time-average over 5000 MD time-steps (10.13 ps). The noisiness of this quantity is also the reason for the large error bars in the figure. The two peaks in the rate correspond to the times that the first evaporation event occur for the two different potentials.

\section{Discussion} 

An arising question is whether the above process is enough to initiate self-sustaining plasma.
In a recent work \cite{ArcPIC_1d,arcPIC}, PIC simulations showed that plasma can build up in the vicinity of an intensively field emitting spot.
This happens if the cathode, emits not only electrons, but also neutral Cu atoms at a rate of at least 0.015 neutrals per emitted electron.
However, the physical processes that can lead to the emission of neutrals along with electrons was not fully understood. 

We define the average evaporation rate $\langle r_{Cu} \rangle$ as the total number of atoms detached from the tip (isolated or clustered) over the total time interval between the first and the last evaporation events. 
Our calculations give $\langle r_{Cu} \rangle = 0.07 \pm 0.01$ atoms/fs with the total number of evaporated atoms being $N_{ev} = 9222 \pm 765$. 
If we divide $r_{Cu}$ by the average emitted electron current during the same period $\langle I \rangle = 2.76 \pm 0.15$ e/fs, we obtain the mean evaporation rate per emitted electron $\langle r_{Cu/e} \rangle = 0.025 \pm 0.003$ atoms/e.
This rate is strikingly close to the values estimated in \cite{arcPIC} and even exceeds the reported minimum of $ r_{Cu/e} = 0.015 \textrm{atoms/e}$ required to ignite plasma.

The consistency of our results with the previous independent simulations that used a different method (PIC) indicates that there is an intrinsic mechanism able to supply neutral atoms sufficient to ignite a vacuum arc near the cathode surface.
Our calculations based on the state-of-the-art understanding of the electric field--material interactions suggest that an external source of neutral atoms might not be necessary to ignite the plasma; the latter can also be fed through self-evaporation of the surface atoms, in a process much less violent than the "explosive emission" conventionally assumed \cite{Anders,Mesyats_Ecton,Mesyats2005}.

Another important question is what are the prerequisites to initiate thermal runaway.
The latter is a complicated process that depends on various initial configuration parameters, namely material, geometry, applied field and voltage.
A full analysis of all these parameters is out of the scope of this work and will be given in a forthcoming publication. 
However, as a general comment, two are the main contributing factors to the initiation of the thermal runaway: melting and force.

This means that the tip height and the current density have to be sufficient to cause melting at the apex region.
Here we have to stress out that the height of the tip itself plays a significant role for the heating of the tip, not only because it increases the field enhancement and thus the emitted current, but also because it decreases the heat dissipation, thus producing higher temperatures at the apex for the same deposited heat.

\begin{table}[h]
	\centering
	\caption{Maximum steady-state temperature for tips of various heights. The applied field is adjusted so that the average current density is $J_{avg} = 2 \pm 0.05 \times 10^{12}\textrm{Am}^{-2}$.}
	\label{tab:tab1}	
	\begin{tabular*}{0.5\linewidth}{l c r}
		\hline \hline 
		height [nm] & & $T_{max}$ [K]	\\
		\hline 
		63.1 & & 1045 \\
		73.1 & & 1067 \\
		83.1 & & 1099 \\
		93.1 & & 1135 \\
		103.1 & &  1152 \\
		\hline
	\end{tabular*}
\end{table} 

This is demonstrated in table 1 where we show the maximum steady-state temperature (constant shape -- no MD simulation) for tips of  various heights. The shape of the upper part of the emitter is the same as in figure \ref{fig:resheat}. The lower "extension" part is adjusted to the given total height. The applied field is adjusted so that the mean current density $J_{avg}$ is kept constant at $2 \times 10^{12}\textrm{Am}^{-2}$. 
We see that although the current density is the same, the tip height itself affects the maximum temperature.

In general, the current density required to produce sufficient heating to reach melting temperatures is of the order of $10^{12} \textrm{A}/\textrm{m}^2$, in agreement with the experimental results of Dyke et. al \cite{Dyke1953I,Dyke1953Arc}. Moreover, the importance of tip melting prior to an arc has been observed also experimentally by Batrakov et. al. \cite{Batrakov_melt}.

On the other hand, the balance between the field-induced forces and the surface stress plays also a significant role.
For the simulations we presented here, a minimum local field at the apex of about 10-12GV/m is necessary.
In view of the above, it would be safe to assume that tips with a height of at least several tens of nm are required in the case of Cu.
The existence of nano-tips of such size, similar to the one we simulated here, is motivated by experimental observations.
Field emission measurements on flat Cu surfaces, have exhibited high enhancement factors of the order 20-100 \cite{CERN2004}. 
Our previous calculations \cite{Veske2016, janssonKMC} has shown that Cu nano-tips that are smaller than 1-2nm in radius cannot be stable. 
Thus, assuming a minimum tip radius of 1-2nm, the tips involved in arc initiation should have a height of at least several tens of nm, which is the main motivation for the geometry chosen to be simulated here.

\section{Conclusions}

In conclusion, we have simulated the thermal and shape evolution of intensively electron emitting Cu nano-tips by means of a multi-scale atomistic model.
Our results reveal a thermal runaway process initiated by the field-induced forces acting on the molten apex of a nano-tip.
The thermal runaway leads to evaporation of metal fractions in the form of either atoms or nano-clusters, at a rate that exceeds the minimum needed to ignite plasma.
This is a self-sufficient process, which does note depend on an external source of neutral atoms.
Thus we show that the onset of a vacuum arc in ultra high vacuum is an intrinsic response of a metal surface to the applied high electric field.

\section*{Acknowledgements}

The current study was supported by the Academy of Finland project AMELIS (grant No. 1269696), CERN CLIC K-contract (No. 47207461), Estonian Research Council Grants PUT 57 and PUT 1372 and the national scholarship program Kristjan Jaak, which is funded and managed by the Archimedes Foundation in collaboration with the Ministry of Education and Research of Estonia. We also acknowledge grants of computer capacity from the Finnish Grid and
Cloud Infrastructure (persistent identifier urn:nbn:fi:research-infras-2016072533).

\pagebreak

\appendix

\section{Simulation flowchart}
\label{sec:flowchart}
\renewcommand\thefigure{\thesection.\arabic{figure}} 
\setcounter{figure}{0}  

\begin{figure}[!h]
	\centering
    \includegraphics[width=\linewidth]{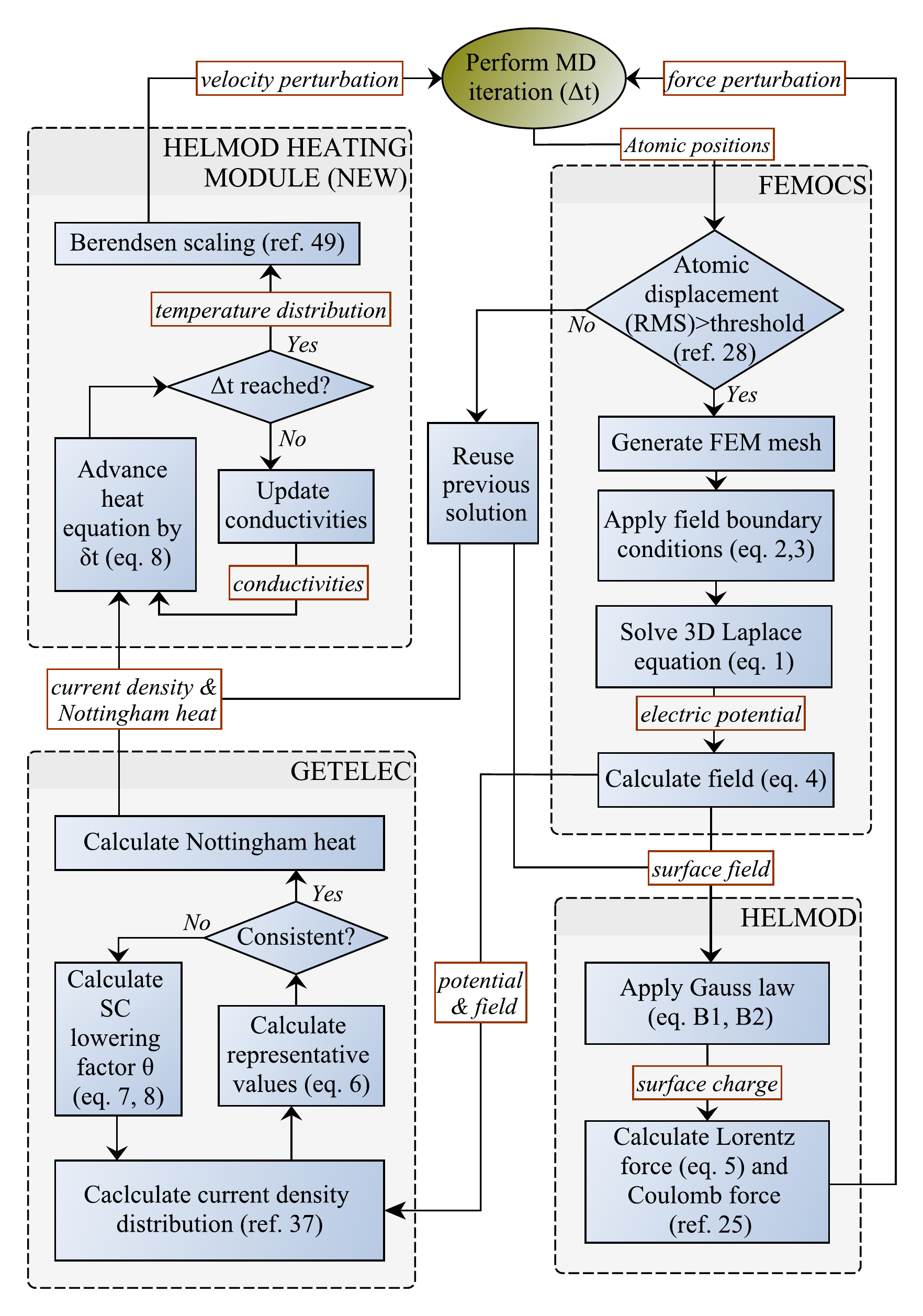}
    \caption{Flowchart of the simulation algorithm.}
    \label{fig:flowchart}
\end{figure} 

\section{Field-induced charge calculation} \label{sec:charges}
\renewcommand\thefigure{\thesection.\arabic{figure}} 
\setcounter{figure}{0}  

In order to calculate the charges and forces on the surface atoms, we use an extension of our previously developed molecular dynamics -- electrodynamics method HELMOD \cite{Djurabekova2011}.
We build a rectangular grid on the Molecular Dynamics (MD) simulation box so that every atom belongs to a certain grid cell. Since the atoms are in an FCC crystal (at least when the temperature is still low), there are numerous grid cells (roughly half) that contain no atoms.
Furthermore, when the temperatures approach or exceed the melting point there might be some cells (a small percentage) containing more than one atoms, but this does not affect the functionality of our method.

\begin{figure}[htbp]
	\centering
    \includegraphics[width=\linewidth]{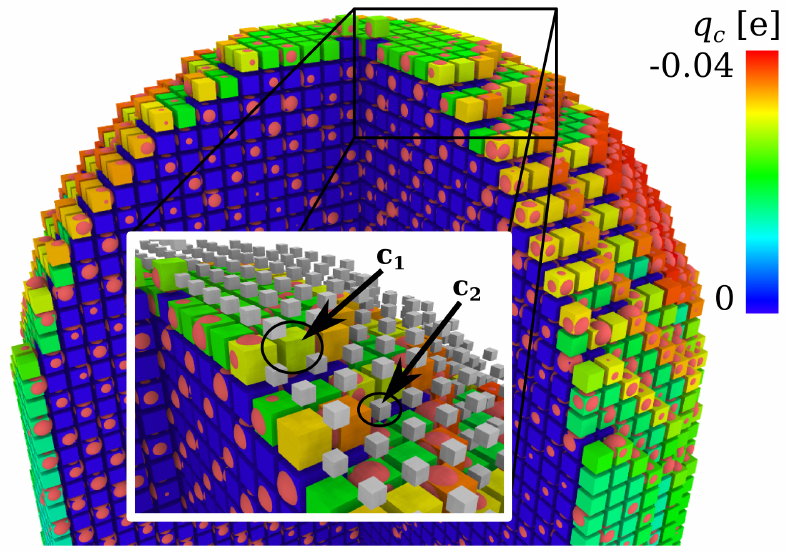}
    \caption{Atoms on the tip apex (maroon spheres) after a relaxation of 0.5ps along with the corresponding material-designated cubic grid cells. The color coding on the cells corresponds to the charge $q_c$ assigned to each cell. The inset demonstrates a closer view in the box area, with the grey small cubes being vacuum-designated cells.}
    \label{fig:grid}
\end{figure} 

When each atom is assigned to a certain cell, the cells are separated into material and vacuum domains. Figure \ref{fig:grid} illustrates this separation. The cubes correspond to the cells of the material domain, for the initial configuration of the simulated tip after a relaxation of 0.5ps. The maroon spheres correspond to the atomic positions. In the inset a closer view of the area in the box is demonstrated, with the smaller grey cubes corresponding to some of the cells of the vacuum domain. 

The separation is done in the following way. If a cell contains an atom or has a number of first nearest neighbours containing an atom $N \geq 4$, then it is designated as material (e.g. cell $c_1$ in the inset). If it does not contain an atom and it has $N \leq 3$ it is designated as vacuum (e.g. cell $c_2$ in the inset). 

Then the surface charges are calculated in every cell face separating a vacuum cell from a material cell according to the Gauss law 
\begin{equation}
	\label{eq:Gauss}
	q_f = \vec{E}_f \cdot \hat{n}_f A_f
\end{equation}
where $q_f$ is the surface charge on each cubic face, $\vec{E}_f$ is the local electric field in the center of the face, and $\hat{n}_f$ is the unit vector normal to the cubic surface.
The electric field distribution has already been calculated by solving the Laplace equation using the recently developed Finite Element Method (FEM) tool FEMOCS \cite{VekseDynamic_arxiv}.
Then the charge on each surface cubic cell is calculated as
\begin{equation}
	\label{eq:Qatom}
	q_c = \sum_{f_{MV}} q_f
\end{equation}
where the summation is done over all the faces of the cell that are in contact with a vacuum one.
The charges $q_c$ are shown in the color coding of the cells in figure \ref{fig:grid}. Note that inner cubic cells that have not a common face with vacuum cells have $q_c=0$.

Since we want to calculate the inter-atomic Coulomb interactions and the field-induced Lorentz forces, the charge on each atom $q_i$ needs to be calculated.
For this, all the grid-point charges have to be assigned to atoms while the total charge is conserved.
This is done in the following way. If a grid point contains one or more atoms, then all its charge $q_c$ is assigned to those atoms (divided equally).
In the vast majority of the cases there is only one atom in each grid cell.
On the other hand, if a grid cell does not contain any atoms, its charge is distributed equally to the atoms belonging to its first-neighbouring cells. 

\section{One-dimensional heat equation} \label{sec:1dheat}

Since the length of the simulated structures is dominant, let us integrate eq. \eqref{eq:3Dheat} on an infinitesimal "slice" $\Omega$, between $z \textrm{ and } z + \delta z$ with width $\delta z$, cross-sectional area $A(z)$ and volume $\delta V = A(z) \delta z$
\begin{equation}
	\label{eq:IntegHeat}
	C_V \frac{\partial }{\partial t} \int_{\Omega}~TdV = \int_{\Omega}~\nabla \left(\kappa \nabla T \right) dV + \int_{\Omega}~pdV \textrm{.}
\end{equation} 

The integral of the left hand side equals to $\langle T \rangle \delta V$ where $\langle T \rangle$ denotes the mean temperature on $\Omega$.
The second term in the right hand side equals to the total heat $\langle p \rangle \delta V$ deposited on the slice.
Using the above notation and Gauss's divergence theorem we can rewrite eq. \eqref{eq:IntegHeat} as
\begin{equation}
	\label{eq:SurHeat}
	C_V \frac{\partial \langle T \rangle}{\partial t} \delta V = \oiint_{\partial \Omega}~\kappa (\nabla T) \cdot \hat{n}  dS + \langle p \rangle \delta V \textrm{.}
\end{equation} 
where $\hat{n}$ denotes the unit vector perpendicular to the surface around the slice $\partial \Omega$.
This surface has three components.
The two cross-sections $A(z)$ and $A(z+\delta z)$ and the ring around them.
For $\delta z \ll 1$ the contribution of the ring is negligible compared to $A$.
Moreover, since the cross-sectional surfaces are along the $x-y$ plane, we have $\hat{n}  \cdot  \nabla T = \partial T / \partial z$ and the surface integral on the cross-section can be expressed as $A(z)\partial \langle T \rangle / \partial z $. Then eq. \eqref{eq:SurHeat} yields
\begin{equation} 
	\label{eq:DifHeat}
	\begin{split}
		C_V \frac{\partial \langle T \rangle}{\partial t} A(z) \delta z = A(z+\delta z) \kappa(z+ \delta z) \left. \frac{\partial \langle T \rangle}{\partial z} \right |_{z+\delta z} \\
	 	- A(z) \kappa(z) \left. \frac{\partial \langle T \rangle}{\partial z} \right |_{z} + \langle p \rangle A(z) \delta z \textrm{.}
	\end{split}
\end{equation}

Dividing eq. \eqref{eq:DifHeat} by $A(z) \delta z$ and considering that the temperature does not vary significantly in the lateral directions, i.e. $\langle T \rangle \approx T$, yields eq. \eqref{eq:1Dheat}, the 1D heat equation for a variable cross-section and conductivity wire.

The total heat $p$ is given by adding the Joule and the Nottingham heat components. The Nottingham heat is expressed as surface density (in $\textrm{W}/\textrm{m}^2$), but it can also be expressed as a volumetric quantity, localized in the surface region with the help of the Dirac $\delta$ function. The total deposited heat at a given point $\vec{r}$ is
\begin{equation} \label{eq:plocal}
	p(\vec{r}) = \frac{J^2(\vec{r})}{\sigma(\vec{r})} + p_N(\vec{r}_s) \delta(\vec{r} - \vec{r}_s) \textrm{,}	
\end{equation}
where $\sigma$ is the local electric conductivity, and $p_N(\vec{r_s})$ is the Nottingham heat at a surface point $\vec{r_s}$.
Then the mean heat $\langle p \rangle$ on a cross-section is given by integrating \eqref{eq:plocal} over the cross-section A(z).
The surface integral of the second term turns into a line integral due to the $\delta$ function and it yields: 
\begin{equation} \label{eq:pslice}
	\langle p \rangle (z) = \frac{1}{A(z)} \left( \frac{I^2(z)}{\sigma(z)} + \oint_{\partial A(z)}~p_N dl \right) \textrm{.}
\end{equation}
In the above equation $I(z)$ is the total current flowing on the cross-section $z$, and $\partial A(z)$ stands for the closed line surrounding the cross-section at $z$. 

\section{Validation of the one-dimensional heat equation} 
\label{sec:1dvalid}

\renewcommand\thefigure{\thesection.\arabic{figure}} 
\setcounter{figure}{0}  

The 3D FEM model of ref. \cite{Eimre2015} is used to solve the steady-state heat equation on the same FEM mesh produced by FEMOCS. The emitted current and Nottingham heat are in that case is calculated using the standard General Thermal-Field (GTF) \cite{Jensen2006}.
Only for this comparison, we will ignore the nanometric emitter size effects in electron emission and the space charge effect.
Note that originally the FEM model ignored the Nottingham effect. Nevertheless, it was here implemented by adding a Neumann boundary condition to the heat equation, similar to equation (4) of reference \cite{Eimre2015}. The boundary condition is
\begin{equation}
	\hat{n} \cdot \left( \kappa \nabla T \right) = p_N
\end{equation}
where $p_N$ denotes the surface density of the deposited heating power and $\hat{n}$ is the unit vector normal to the material surface.

\begin{figure}[htbp]
	\centering
    \includegraphics[width=\linewidth]{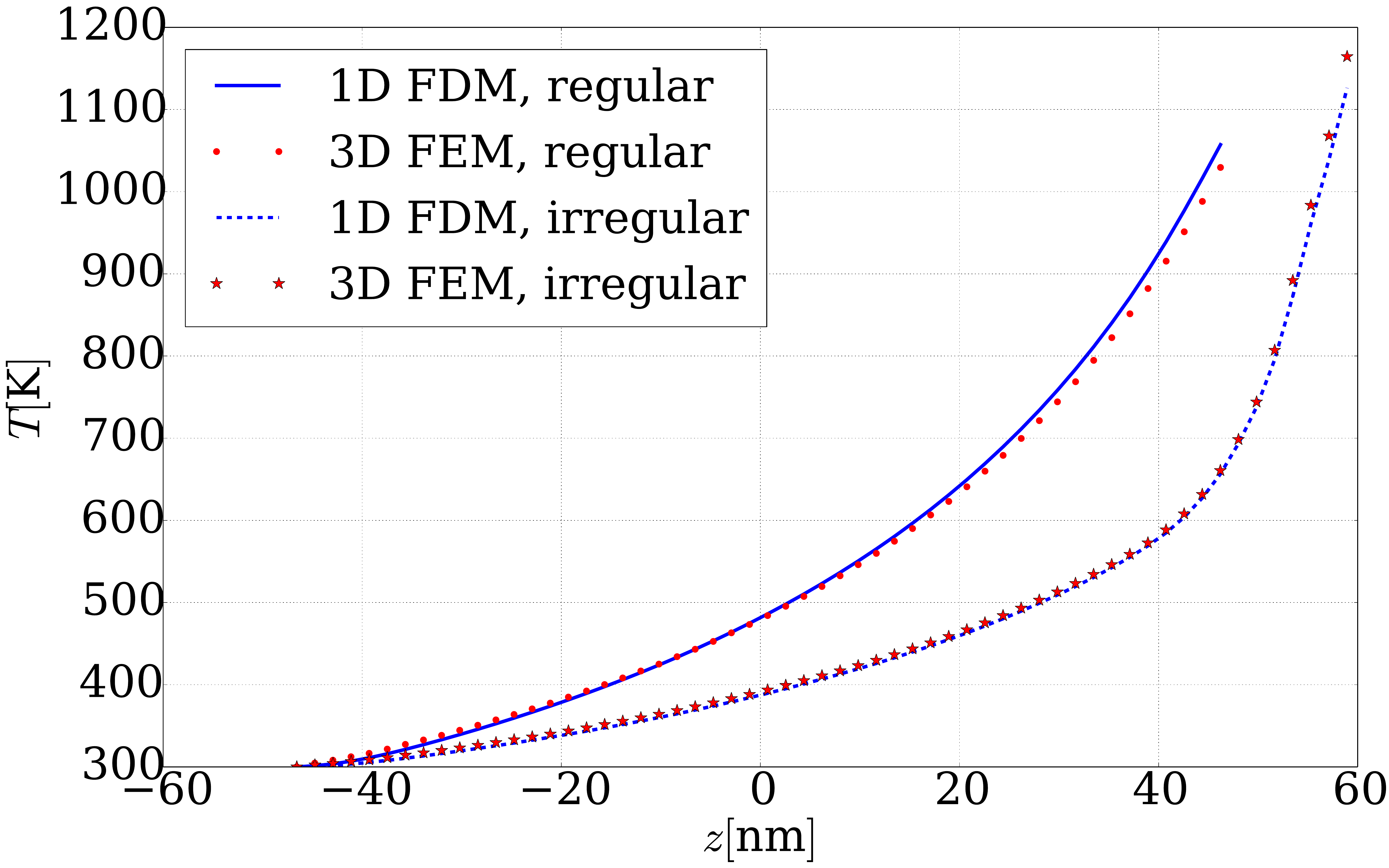}
    \caption{Steady state temperature distribution as calculated by the 1D FDM (blue lines) and the 3D FEM (red markers) models for two different tip geometries and field values. For the regular conical geometry "regular" (solid lines) the applied field was $E_{0}=0.56 \textrm{GV}/\textrm{m}$ and for the oblique geometry "irregular" (dashed lines) $E_0=0.45 \textrm{GV}/\textrm{m}$.}
    \label{fig:tempcomp}
\end{figure}

Figure \ref{fig:tempcomp} shows the comparison between the two models.
The lines correspond to the final steady-state temperature distribution along the simulated tip, as calculated with the 1D model described above.
The markers correspond to the temperature along the vertical axis of symmetry of the tip, as calculated by the 3D FEM model.

The calculation was performed for two different tip geometries, a "regular" and an "irregular". 
The "regular tip" geometry (solid lines and dots) corresponds to the original conical geometry (shown in figure \ref{fig:resheat}) and the one for "irregular tip" one(dashed lines and diamonds) to the deformed shape of frame (b) in figure \ref{fig:frames}.
We see that the simplified 1D model is in excellent agreement with the 3D FEM calculation, even for the irregular geometry.
The markers correspond to the mean value of the temperature for a given slice.
The deviation of the temperature around that value is smaller than the size of the marker in the figure. 

\pagebreak

\bibliography{bibliography/bibliography}

\begin{thebibliography}{58}%
\makeatletter
\providecommand \@ifxundefined [1]{%
 \@ifx{#1\undefined}
}%
\providecommand \@ifnum [1]{%
 \ifnum #1\expandafter \@firstoftwo
 \else \expandafter \@secondoftwo
 \fi
}%
\providecommand \@ifx [1]{%
 \ifx #1\expandafter \@firstoftwo
 \else \expandafter \@secondoftwo
 \fi
}%
\providecommand \natexlab [1]{#1}%
\providecommand \enquote  [1]{``#1''}%
\providecommand \bibnamefont  [1]{#1}%
\providecommand \bibfnamefont [1]{#1}%
\providecommand \citenamefont [1]{#1}%
\providecommand \href@noop [0]{\@secondoftwo}%
\providecommand \href [0]{\begingroup \@sanitize@url \@href}%
\providecommand \@href[1]{\@@startlink{#1}\@@href}%
\providecommand \@@href[1]{\endgroup#1\@@endlink}%
\providecommand \@sanitize@url [0]{\catcode `\\12\catcode `\$12\catcode
  `\&12\catcode `\#12\catcode `\^12\catcode `\_12\catcode `\%12\relax}%
\providecommand \@@startlink[1]{}%
\providecommand \@@endlink[0]{}%
\providecommand \url  [0]{\begingroup\@sanitize@url \@url }%
\providecommand \@url [1]{\endgroup\@href {#1}{\urlprefix }}%
\providecommand \urlprefix  [0]{URL }%
\providecommand \Eprint [0]{\href }%
\providecommand \doibase [0]{http://dx.doi.org/}%
\providecommand \selectlanguage [0]{\@gobble}%
\providecommand \bibinfo  [0]{\@secondoftwo}%
\providecommand \bibfield  [0]{\@secondoftwo}%
\providecommand \translation [1]{[#1]}%
\providecommand \BibitemOpen [0]{}%
\providecommand \bibitemStop [0]{}%
\providecommand \bibitemNoStop [0]{.\EOS\space}%
\providecommand \EOS [0]{\spacefactor3000\relax}%
\providecommand \BibitemShut  [1]{\csname bibitem#1\endcsname}%
\let\auto@bib@innerbib\@empty
\bibitem [{\citenamefont {Egorov}\ and\ \citenamefont
  {Sheshin}(2017)}]{egorov2017field}%
  \BibitemOpen
  \bibfield  {author} {\bibinfo {author} {\bibfnamefont {N.}~\bibnamefont
  {Egorov}}\ and\ \bibinfo {author} {\bibfnamefont {E.}~\bibnamefont
  {Sheshin}},\ }\href@noop {} {\emph {\bibinfo {title} {Field Emission
  Electronics}}}\ (\bibinfo  {publisher} {Springer},\ \bibinfo {year}
  {2017})\BibitemShut {NoStop}%
\bibitem [{\citenamefont {Brown}\ \emph {et~al.}(1985)\citenamefont {Brown},
  \citenamefont {Galvin},\ and\ \citenamefont {MacGill}}]{MEVVA1985}%
  \BibitemOpen
  \bibfield  {author} {\bibinfo {author} {\bibfnamefont {I.~G.}\ \bibnamefont
  {Brown}}, \bibinfo {author} {\bibfnamefont {J.~E.}\ \bibnamefont {Galvin}}, \
  and\ \bibinfo {author} {\bibfnamefont {R.~A.}\ \bibnamefont {MacGill}},\
  }\href {\doibase http://dx.doi.org/10.1063/1.96163} {\bibfield  {journal}
  {\bibinfo  {journal} {Applied Physics Letters}\ }\textbf {\bibinfo {volume}
  {47}},\ \bibinfo {pages} {358} (\bibinfo {year} {1985})}\BibitemShut
  {NoStop}%
\bibitem [{\citenamefont {Anders}(2008)}]{Anders}%
  \BibitemOpen
  \bibfield  {author} {\bibinfo {author} {\bibfnamefont {A.}~\bibnamefont
  {Anders}},\ }\href {\doibase 10.1007/978-0-387-79108-1} {\emph {\bibinfo
  {title} {Cathodic Arcs}}},\ \bibinfo {edition} {1st}\ ed.,\ Vol.~\bibinfo
  {volume} {3}\ (\bibinfo  {publisher} {Springer-Verlag New York},\ \bibinfo
  {year} {2008})\BibitemShut {NoStop}%
\bibitem [{\citenamefont {Hartmann}\ and\ \citenamefont
  {Gundersen}(1988)}]{Anders_PRL88}%
  \BibitemOpen
  \bibfield  {author} {\bibinfo {author} {\bibfnamefont {W.}~\bibnamefont
  {Hartmann}}\ and\ \bibinfo {author} {\bibfnamefont {M.~A.}\ \bibnamefont
  {Gundersen}},\ }\href {\doibase 10.1103/PhysRevLett.60.2371} {\bibfield
  {journal} {\bibinfo  {journal} {Physical Review Letters}\ }\textbf {\bibinfo
  {volume} {60}},\ \bibinfo {pages} {2371} (\bibinfo {year}
  {1988})}\BibitemShut {NoStop}%
\bibitem [{\citenamefont {Dyke}\ and\ \citenamefont
  {Trolan}(1953)}]{Dyke1953Arc}%
  \BibitemOpen
  \bibfield  {author} {\bibinfo {author} {\bibfnamefont {W.}~\bibnamefont
  {Dyke}}\ and\ \bibinfo {author} {\bibfnamefont {J.}~\bibnamefont {Trolan}},\
  }\href {\doibase 10.1103/PhysRev.89.799} {\bibfield  {journal} {\bibinfo
  {journal} {Physical Review}\ }\textbf {\bibinfo {volume} {89}},\ \bibinfo
  {pages} {799} (\bibinfo {year} {1953})}\BibitemShut {NoStop}%
\bibitem [{\citenamefont {Descoeudres}\ \emph
  {et~al.}(2009{\natexlab{a}})\citenamefont {Descoeudres}, \citenamefont
  {Ramsvik}, \citenamefont {Calatroni}, \citenamefont {Taborelli},\ and\
  \citenamefont {Wuensch}}]{Descoeudres2009_dc}%
  \BibitemOpen
  \bibfield  {author} {\bibinfo {author} {\bibfnamefont {A.}~\bibnamefont
  {Descoeudres}}, \bibinfo {author} {\bibfnamefont {T.}~\bibnamefont
  {Ramsvik}}, \bibinfo {author} {\bibfnamefont {S.}~\bibnamefont {Calatroni}},
  \bibinfo {author} {\bibfnamefont {M.}~\bibnamefont {Taborelli}}, \ and\
  \bibinfo {author} {\bibfnamefont {W.}~\bibnamefont {Wuensch}},\ }\href
  {\doibase 10.1103/PhysRevSTAB.12.032001} {\bibfield  {journal} {\bibinfo
  {journal} {Physical Review Special Topics - Accelerators and Beams}\ }\textbf
  {\bibinfo {volume} {12}},\ \bibinfo {pages} {032001} (\bibinfo {year}
  {2009}{\natexlab{a}})}\BibitemShut {NoStop}%
\bibitem [{\citenamefont {McCracken}(1980)}]{McCracken1980}%
  \BibitemOpen
  \bibfield  {author} {\bibinfo {author} {\bibfnamefont {G.}~\bibnamefont
  {McCracken}},\ }\href {\doibase https://doi.org/10.1016/0022-3115(80)90299-8}
  {\bibfield  {journal} {\bibinfo  {journal} {Journal of Nuclear Materials}\
  }\textbf {\bibinfo {volume} {93}},\ \bibinfo {pages} {3 } (\bibinfo {year}
  {1980})}\BibitemShut {NoStop}%
\bibitem [{\citenamefont {Slade}(2007)}]{slade2007}%
  \BibitemOpen
  \bibfield  {author} {\bibinfo {author} {\bibfnamefont {P.}~\bibnamefont
  {Slade}},\ }\href {https://books.google.fi/books?id=uJT4mENgMbQC} {\emph
  {\bibinfo {title} {The Vacuum Interrupter: Theory, Design, and
  Application}}}\ (\bibinfo  {publisher} {CRC Press},\ \bibinfo {year}
  {2007})\BibitemShut {NoStop}%
\bibitem [{\citenamefont {Aicheler}\ \emph {et~al.}(2012)\citenamefont
  {Aicheler}, \citenamefont {Burrows}, \citenamefont {Draper}, \citenamefont
  {Garvey}, \citenamefont {Lebrun}, \citenamefont {Peach}, \citenamefont
  {Phinney}, \citenamefont {Schmickler}, \citenamefont {Schulte},\ and\
  \citenamefont {Tog}}]{clic}%
  \BibitemOpen
  \bibinfo {editor} {\bibfnamefont {M.}~\bibnamefont {Aicheler}}, \bibinfo
  {editor} {\bibfnamefont {P.}~\bibnamefont {Burrows}}, \bibinfo {editor}
  {\bibfnamefont {M.}~\bibnamefont {Draper}}, \bibinfo {editor} {\bibfnamefont
  {T.}~\bibnamefont {Garvey}}, \bibinfo {editor} {\bibfnamefont
  {P.}~\bibnamefont {Lebrun}}, \bibinfo {editor} {\bibfnamefont
  {K.}~\bibnamefont {Peach}}, \bibinfo {editor} {\bibfnamefont
  {N.}~\bibnamefont {Phinney}}, \bibinfo {editor} {\bibfnamefont
  {H.}~\bibnamefont {Schmickler}}, \bibinfo {editor} {\bibfnamefont
  {D.}~\bibnamefont {Schulte}}, \ and\ \bibinfo {editor} {\bibfnamefont
  {N.}~\bibnamefont {Tog}},\ eds.,\ \href@noop {} {\emph {\bibinfo {title} {{A
  Multi-TeV linear collider based on CLIC technology: CLIC Conceptual Design
  Report}}}}\ (\bibinfo  {publisher} {CERN},\ \bibinfo {year}
  {2012})\BibitemShut {NoStop}%
\bibitem [{\citenamefont {Dyke}\ \emph {et~al.}(1953)\citenamefont {Dyke},
  \citenamefont {Trolan}, \citenamefont {Martin},\ and\ \citenamefont
  {Barbour}}]{Dyke1953I}%
  \BibitemOpen
  \bibfield  {author} {\bibinfo {author} {\bibfnamefont {W.}~\bibnamefont
  {Dyke}}, \bibinfo {author} {\bibfnamefont {J.}~\bibnamefont {Trolan}},
  \bibinfo {author} {\bibfnamefont {E.}~\bibnamefont {Martin}}, \ and\ \bibinfo
  {author} {\bibfnamefont {J.}~\bibnamefont {Barbour}},\ }\href {\doibase
  10.1103/PhysRev.91.1043} {\bibfield  {journal} {\bibinfo  {journal} {Physical
  Review}\ }\textbf {\bibinfo {volume} {91}},\ \bibinfo {pages} {1043}
  (\bibinfo {year} {1953})}\BibitemShut {NoStop}%
\bibitem [{\citenamefont {Dolan}\ \emph {et~al.}(1953)\citenamefont {Dolan},
  \citenamefont {Dyke},\ and\ \citenamefont {Trolan}}]{DolanII}%
  \BibitemOpen
  \bibfield  {author} {\bibinfo {author} {\bibfnamefont {W.}~\bibnamefont
  {Dolan}}, \bibinfo {author} {\bibfnamefont {W.}~\bibnamefont {Dyke}}, \ and\
  \bibinfo {author} {\bibfnamefont {J.}~\bibnamefont {Trolan}},\ }\href
  {\doibase 10.1103/PhysRev.91.1054} {\bibfield  {journal} {\bibinfo  {journal}
  {Physical Review}\ }\textbf {\bibinfo {volume} {91}},\ \bibinfo {pages}
  {1054} (\bibinfo {year} {1953})}\BibitemShut {NoStop}%
\bibitem [{\citenamefont {Anders}\ \emph {et~al.}(1993)\citenamefont {Anders},
  \citenamefont {Anders},\ and\ \citenamefont {Gundersen}}]{Anders_PRL93}%
  \BibitemOpen
  \bibfield  {author} {\bibinfo {author} {\bibfnamefont {A.}~\bibnamefont
  {Anders}}, \bibinfo {author} {\bibfnamefont {S.}~\bibnamefont {Anders}}, \
  and\ \bibinfo {author} {\bibfnamefont {M.~A.}\ \bibnamefont {Gundersen}},\
  }\href {\doibase 10.1103/PhysRevLett.71.364} {\bibfield  {journal} {\bibinfo
  {journal} {Physical Review Letters}\ }\textbf {\bibinfo {volume} {71}},\
  \bibinfo {pages} {364} (\bibinfo {year} {1993})}\BibitemShut {NoStop}%
\bibitem [{\citenamefont {Latham}(1995)}]{latham1995}%
  \BibitemOpen
  \bibfield  {author} {\bibinfo {author} {\bibfnamefont {R.~V.}\ \bibnamefont
  {Latham}},\ }\href@noop {} {\emph {\bibinfo {title} {High voltage vacuum
  insulation: Basic concepts and technological practice}}}\ (\bibinfo
  {publisher} {Elsevier},\ \bibinfo {year} {1995})\BibitemShut {NoStop}%
\bibitem [{\citenamefont {Descoeudres}\ \emph
  {et~al.}(2009{\natexlab{b}})\citenamefont {Descoeudres}, \citenamefont
  {Levinsen}, \citenamefont {Calatroni}, \citenamefont {Taborelli},\ and\
  \citenamefont {Wuensch}}]{Descoeudres2009}%
  \BibitemOpen
  \bibfield  {author} {\bibinfo {author} {\bibfnamefont {A.}~\bibnamefont
  {Descoeudres}}, \bibinfo {author} {\bibfnamefont {Y.}~\bibnamefont
  {Levinsen}}, \bibinfo {author} {\bibfnamefont {S.}~\bibnamefont {Calatroni}},
  \bibinfo {author} {\bibfnamefont {M.}~\bibnamefont {Taborelli}}, \ and\
  \bibinfo {author} {\bibfnamefont {W.}~\bibnamefont {Wuensch}},\ }\href
  {\doibase 10.1103/PhysRevSTAB.12.092001} {\bibfield  {journal} {\bibinfo
  {journal} {Physical Review Special Topics - Accelerators and Beams}\ }\textbf
  {\bibinfo {volume} {12}},\ \bibinfo {pages} {092001} (\bibinfo {year}
  {2009}{\natexlab{b}})}\BibitemShut {NoStop}%
\bibitem [{\citenamefont {Degiovanni}\ \emph {et~al.}(2016)\citenamefont
  {Degiovanni}, \citenamefont {Wuensch},\ and\ \citenamefont
  {Giner~Navarro}}]{Degiovani_conditioning}%
  \BibitemOpen
  \bibfield  {author} {\bibinfo {author} {\bibfnamefont {A.}~\bibnamefont
  {Degiovanni}}, \bibinfo {author} {\bibfnamefont {W.}~\bibnamefont {Wuensch}},
  \ and\ \bibinfo {author} {\bibfnamefont {J.}~\bibnamefont {Giner~Navarro}},\
  }\href {\doibase 10.1103/PhysRevAccelBeams.19.032001} {\bibfield  {journal}
  {\bibinfo  {journal} {Physical Review Special Topics - Accelerators and
  Beams}\ }\textbf {\bibinfo {volume} {19}},\ \bibinfo {pages} {032001}
  (\bibinfo {year} {2016})}\BibitemShut {NoStop}%
\bibitem [{\citenamefont {Mesyats}(1995)}]{Mesyats_Ecton}%
  \BibitemOpen
  \bibfield  {author} {\bibinfo {author} {\bibfnamefont {G.~A.}\ \bibnamefont
  {Mesyats}},\ }\href {\doibase 10.1109/27.476469} {\bibfield  {journal}
  {\bibinfo  {journal} {IEEE Transansactions on Plasma Science}\ }\textbf
  {\bibinfo {volume} {23}},\ \bibinfo {pages} {879} (\bibinfo {year}
  {1995})}\BibitemShut {NoStop}%
\bibitem [{\citenamefont {Mesyats}(1993)}]{mesyats1993ectons}%
  \BibitemOpen
  \bibfield  {author} {\bibinfo {author} {\bibfnamefont {G.}~\bibnamefont
  {Mesyats}},\ }\href@noop {} {\bibfield  {journal} {\bibinfo  {journal} {JETP
  LETTERS}\ }\textbf {\bibinfo {volume} {57}},\ \bibinfo {pages} {95} (\bibinfo
  {year} {1993})}\BibitemShut {NoStop}%
\bibitem [{\citenamefont {Mesyats}(2005)}]{Mesyats2005}%
  \BibitemOpen
  \bibfield  {author} {\bibinfo {author} {\bibfnamefont {G.~A.}\ \bibnamefont
  {Mesyats}},\ }\href {http://stacks.iop.org/0741-3335/47/i=5A/a=010}
  {\bibfield  {journal} {\bibinfo  {journal} {Plasma Physics and Controlled
  Fusion}\ }\textbf {\bibinfo {volume} {47}},\ \bibinfo {pages} {A109}
  (\bibinfo {year} {2005})}\BibitemShut {NoStop}%
\bibitem [{\citenamefont {Krohn}\ and\ \citenamefont
  {Ringo}(1975)}]{krohn1975ion}%
  \BibitemOpen
  \bibfield  {author} {\bibinfo {author} {\bibfnamefont {V.}~\bibnamefont
  {Krohn}}\ and\ \bibinfo {author} {\bibfnamefont {G.}~\bibnamefont {Ringo}},\
  }\href@noop {} {\bibfield  {journal} {\bibinfo  {journal} {Applied Physics
  Letters}\ }\textbf {\bibinfo {volume} {27}},\ \bibinfo {pages} {479}
  (\bibinfo {year} {1975})}\BibitemShut {NoStop}%
\bibitem [{\citenamefont {Wagner}(1982)}]{Wagner1982hydrodynamics}%
  \BibitemOpen
  \bibfield  {author} {\bibinfo {author} {\bibfnamefont {A.}~\bibnamefont
  {Wagner}},\ }\href {\doibase 10.1063/1.93100} {\bibfield  {journal} {\bibinfo
   {journal} {Applied Physics Letters}\ }\textbf {\bibinfo {volume} {40}},\
  \bibinfo {pages} {440} (\bibinfo {year} {1982})}\BibitemShut {NoStop}%
\bibitem [{\citenamefont {Swanson}(1983)}]{Swanson_LMIS}%
  \BibitemOpen
  \bibfield  {author} {\bibinfo {author} {\bibfnamefont {L.}~\bibnamefont
  {Swanson}},\ }\href {\doibase http://dx.doi.org/10.1016/0167-5087(83)91005-0}
  {\bibfield  {journal} {\bibinfo  {journal} {Nuclear Instruments and Methods
  in Physics Research}\ }\textbf {\bibinfo {volume} {218}},\ \bibinfo {pages}
  {347 } (\bibinfo {year} {1983})}\BibitemShut {NoStop}%
\bibitem [{\citenamefont {Timko}\ \emph {et~al.}(2011)\citenamefont {Timko},
  \citenamefont {Matyash}, \citenamefont {Schneider}, \citenamefont
  {Djurabekova}, \citenamefont {Nordlund}, \citenamefont {Hansen},
  \citenamefont {Descoeudres}, \citenamefont {Kovermann}, \citenamefont
  {Grudiev}, \citenamefont {Wuensch}, \citenamefont {Calatroni},\ and\
  \citenamefont {Taborelli}}]{ArcPIC_1d}%
  \BibitemOpen
  \bibfield  {author} {\bibinfo {author} {\bibfnamefont {H.}~\bibnamefont
  {Timko}}, \bibinfo {author} {\bibfnamefont {K.}~\bibnamefont {Matyash}},
  \bibinfo {author} {\bibfnamefont {R.}~\bibnamefont {Schneider}}, \bibinfo
  {author} {\bibfnamefont {F.}~\bibnamefont {Djurabekova}}, \bibinfo {author}
  {\bibfnamefont {K.}~\bibnamefont {Nordlund}}, \bibinfo {author}
  {\bibfnamefont {A.}~\bibnamefont {Hansen}}, \bibinfo {author} {\bibfnamefont
  {A.}~\bibnamefont {Descoeudres}}, \bibinfo {author} {\bibfnamefont
  {J.}~\bibnamefont {Kovermann}}, \bibinfo {author} {\bibfnamefont
  {A.}~\bibnamefont {Grudiev}}, \bibinfo {author} {\bibfnamefont
  {W.}~\bibnamefont {Wuensch}}, \bibinfo {author} {\bibfnamefont
  {S.}~\bibnamefont {Calatroni}}, \ and\ \bibinfo {author} {\bibfnamefont
  {M.}~\bibnamefont {Taborelli}},\ }\href {\doibase 10.1002/ctpp.201000504}
  {\bibfield  {journal} {\bibinfo  {journal} {Contributions to Plasma Physics}\
  }\textbf {\bibinfo {volume} {51}},\ \bibinfo {pages} {5} (\bibinfo {year}
  {2011})}\BibitemShut {NoStop}%
\bibitem [{\citenamefont {Timko}\ \emph {et~al.}(2015)\citenamefont {Timko},
  \citenamefont {Ness~Sjobak}, \citenamefont {Mether}, \citenamefont
  {Calatroni}, \citenamefont {Djurabekova}, \citenamefont {Matyash},
  \citenamefont {Nordlund}, \citenamefont {Schneider},\ and\ \citenamefont
  {Wuensch}}]{arcPIC}%
  \BibitemOpen
  \bibfield  {author} {\bibinfo {author} {\bibfnamefont {H.}~\bibnamefont
  {Timko}}, \bibinfo {author} {\bibfnamefont {K.}~\bibnamefont {Ness~Sjobak}},
  \bibinfo {author} {\bibfnamefont {L.}~\bibnamefont {Mether}}, \bibinfo
  {author} {\bibfnamefont {S.}~\bibnamefont {Calatroni}}, \bibinfo {author}
  {\bibfnamefont {F.}~\bibnamefont {Djurabekova}}, \bibinfo {author}
  {\bibfnamefont {K.}~\bibnamefont {Matyash}}, \bibinfo {author} {\bibfnamefont
  {K.}~\bibnamefont {Nordlund}}, \bibinfo {author} {\bibfnamefont
  {R.}~\bibnamefont {Schneider}}, \ and\ \bibinfo {author} {\bibfnamefont
  {W.}~\bibnamefont {Wuensch}},\ }\href {\doibase 10.1002/ctpp.201400069}
  {\bibfield  {journal} {\bibinfo  {journal} {Contributions to Plasma Physics}\
  }\textbf {\bibinfo {volume} {55}},\ \bibinfo {pages} {299} (\bibinfo {year}
  {2015})}\BibitemShut {NoStop}%
\bibitem [{\citenamefont {Nottingham}(1936)}]{Nottingham}%
  \BibitemOpen
  \bibfield  {author} {\bibinfo {author} {\bibfnamefont {W.~B.}\ \bibnamefont
  {Nottingham}},\ }\href {\doibase 10.1103/PhysRev.49.78} {\bibfield  {journal}
  {\bibinfo  {journal} {Physical Review}\ }\textbf {\bibinfo {volume} {49}},\
  \bibinfo {pages} {78} (\bibinfo {year} {1936})}\BibitemShut {NoStop}%
\bibitem [{\citenamefont {Charbonnier}\ \emph {et~al.}(1964)\citenamefont
  {Charbonnier}, \citenamefont {Strayer}, \citenamefont {Swanson},\ and\
  \citenamefont {Martin}}]{NottingCharb}%
  \BibitemOpen
  \bibfield  {author} {\bibinfo {author} {\bibfnamefont {F.~M.}\ \bibnamefont
  {Charbonnier}}, \bibinfo {author} {\bibfnamefont {R.~W.}\ \bibnamefont
  {Strayer}}, \bibinfo {author} {\bibfnamefont {L.~W.}\ \bibnamefont
  {Swanson}}, \ and\ \bibinfo {author} {\bibfnamefont {E.~E.}\ \bibnamefont
  {Martin}},\ }\href {\doibase 10.1103/PhysRevLett.13.397} {\bibfield
  {journal} {\bibinfo  {journal} {Physical Review Letters}\ }\textbf {\bibinfo
  {volume} {13}},\ \bibinfo {pages} {397} (\bibinfo {year} {1964})}\BibitemShut
  {NoStop}%
\bibitem [{\citenamefont {Djurabekova}\ \emph {et~al.}(2011)\citenamefont
  {Djurabekova}, \citenamefont {Parviainen}, \citenamefont {Pohjonen},\ and\
  \citenamefont {Nordlund}}]{Djurabekova2011}%
  \BibitemOpen
  \bibfield  {author} {\bibinfo {author} {\bibfnamefont {F.}~\bibnamefont
  {Djurabekova}}, \bibinfo {author} {\bibfnamefont {S.}~\bibnamefont
  {Parviainen}}, \bibinfo {author} {\bibfnamefont {A.}~\bibnamefont
  {Pohjonen}}, \ and\ \bibinfo {author} {\bibfnamefont {K.}~\bibnamefont
  {Nordlund}},\ }\href {\doibase 10.1103/PhysRevE.83.026704} {\bibfield
  {journal} {\bibinfo  {journal} {Physical Review E}\ }\textbf {\bibinfo
  {volume} {83}},\ \bibinfo {pages} {026704} (\bibinfo {year}
  {2011})}\BibitemShut {NoStop}%
\bibitem [{\citenamefont {Veske}\ \emph
  {et~al.}(2016{\natexlab{a}})\citenamefont {Veske}, \citenamefont
  {Kyritsakis}, \citenamefont {Djurabekova}, \citenamefont {Aare},
  \citenamefont {Eimre},\ and\ \citenamefont {Zadin}}]{FemocsIVNC}%
  \BibitemOpen
  \bibfield  {author} {\bibinfo {author} {\bibfnamefont {M.}~\bibnamefont
  {Veske}}, \bibinfo {author} {\bibfnamefont {A.}~\bibnamefont {Kyritsakis}},
  \bibinfo {author} {\bibfnamefont {F.}~\bibnamefont {Djurabekova}}, \bibinfo
  {author} {\bibfnamefont {R.}~\bibnamefont {Aare}}, \bibinfo {author}
  {\bibfnamefont {K.}~\bibnamefont {Eimre}}, \ and\ \bibinfo {author}
  {\bibfnamefont {V.}~\bibnamefont {Zadin}},\ }in\ \href {\doibase
  10.1109/IVNC.2016.7551501} {\emph {\bibinfo {booktitle} {2016 29th
  International Vacuum Nanoelectronics Conference (IVNC)}}}\ (\bibinfo {year}
  {2016})\ pp.\ \bibinfo {pages} {1--2}\BibitemShut {NoStop}%
\bibitem [{\citenamefont {{Veske}}\ \emph {et~al.}(2017)\citenamefont
  {{Veske}}, \citenamefont {{Kyritsakis}}, \citenamefont {{Eimre}},
  \citenamefont {{Zadin}}, \citenamefont {{Aabloo}},\ and\ \citenamefont
  {{Djurabekova}}}]{VekseDynamic_arxiv}%
  \BibitemOpen
  \bibfield  {author} {\bibinfo {author} {\bibfnamefont {M.}~\bibnamefont
  {{Veske}}}, \bibinfo {author} {\bibfnamefont {A.}~\bibnamefont
  {{Kyritsakis}}}, \bibinfo {author} {\bibfnamefont {K.}~\bibnamefont
  {{Eimre}}}, \bibinfo {author} {\bibfnamefont {V.}~\bibnamefont {{Zadin}}},
  \bibinfo {author} {\bibfnamefont {A.}~\bibnamefont {{Aabloo}}}, \ and\
  \bibinfo {author} {\bibfnamefont {F.}~\bibnamefont {{Djurabekova}}},\
  }\href@noop {} {\bibfield  {journal} {\bibinfo  {journal} {ArXiv e-prints}\ }
  (\bibinfo {year} {2017})},\ \Eprint {http://arxiv.org/abs/1706.09661v2}
  {arXiv:1706.09661v2} \BibitemShut {NoStop}%
\bibitem [{\citenamefont {Parviainen}\ \emph {et~al.}(2011)\citenamefont
  {Parviainen}, \citenamefont {Djurabekova}, \citenamefont {Timko},\ and\
  \citenamefont {Nordlund}}]{Parviainen2011}%
  \BibitemOpen
  \bibfield  {author} {\bibinfo {author} {\bibfnamefont {S.}~\bibnamefont
  {Parviainen}}, \bibinfo {author} {\bibfnamefont {F.}~\bibnamefont
  {Djurabekova}}, \bibinfo {author} {\bibfnamefont {H.}~\bibnamefont {Timko}},
  \ and\ \bibinfo {author} {\bibfnamefont {K.}~\bibnamefont {Nordlund}},\
  }\href {\doibase 10.1016/j.commatsci.2011.02.010} {\bibfield  {journal}
  {\bibinfo  {journal} {Computational Materials Science}\ }\textbf {\bibinfo
  {volume} {50}},\ \bibinfo {pages} {2075} (\bibinfo {year}
  {2011})}\BibitemShut {NoStop}%
\bibitem [{\citenamefont {Eimre}\ \emph {et~al.}(2015)\citenamefont {Eimre},
  \citenamefont {Parviainen}, \citenamefont {Aabloo}, \citenamefont
  {Djurabekova},\ and\ \citenamefont {Zadin}}]{Eimre2015}%
  \BibitemOpen
  \bibfield  {author} {\bibinfo {author} {\bibfnamefont {K.}~\bibnamefont
  {Eimre}}, \bibinfo {author} {\bibfnamefont {S.}~\bibnamefont {Parviainen}},
  \bibinfo {author} {\bibfnamefont {A.}~\bibnamefont {Aabloo}}, \bibinfo
  {author} {\bibfnamefont {F.}~\bibnamefont {Djurabekova}}, \ and\ \bibinfo
  {author} {\bibfnamefont {V.}~\bibnamefont {Zadin}},\ }\href {\doibase
  10.1063/1.4926490} {\bibfield  {journal} {\bibinfo  {journal} {Journal of
  Applied Physics}\ }\textbf {\bibinfo {volume} {118}},\ \bibinfo {pages}
  {033303} (\bibinfo {year} {2015})}\BibitemShut {NoStop}%
\bibitem [{\citenamefont {Schottky}(1923)}]{Schottky1923}%
  \BibitemOpen
  \bibfield  {author} {\bibinfo {author} {\bibfnamefont {W.}~\bibnamefont
  {Schottky}},\ }\href {\doibase 10.1007/BF01340034} {\bibfield  {journal}
  {\bibinfo  {journal} {Zeitschrift für Physik}\ }\textbf {\bibinfo {volume}
  {14}},\ \bibinfo {pages} {63} (\bibinfo {year} {1923})}\BibitemShut {NoStop}%
\bibitem [{\citenamefont {Fowler}\ and\ \citenamefont
  {Nordheim}(1928)}]{FN1928}%
  \BibitemOpen
  \bibfield  {author} {\bibinfo {author} {\bibfnamefont {R.~H.}\ \bibnamefont
  {Fowler}}\ and\ \bibinfo {author} {\bibfnamefont {L.}~\bibnamefont
  {Nordheim}},\ }\href {\doibase 10.1098/rspa.1928.0091} {\bibfield  {journal}
  {\bibinfo  {journal} {Proceedings of the Royal Society of London A}\ }\textbf
  {\bibinfo {volume} {119}},\ \bibinfo {pages} {173} (\bibinfo {year}
  {1928})}\BibitemShut {NoStop}%
\bibitem [{\citenamefont {Murphy}\ and\ \citenamefont {Good}(1956)}]{MurphyG}%
  \BibitemOpen
  \bibfield  {author} {\bibinfo {author} {\bibfnamefont {E.~L.}\ \bibnamefont
  {Murphy}}\ and\ \bibinfo {author} {\bibfnamefont {R.~H.}\ \bibnamefont
  {Good}},\ }\href {\doibase 10.1103/PhysRev.102.1464} {\bibfield  {journal}
  {\bibinfo  {journal} {Physical Review}\ }\textbf {\bibinfo {volume} {102}},\
  \bibinfo {pages} {1464} (\bibinfo {year} {1956})}\BibitemShut {NoStop}%
\bibitem [{\citenamefont {Jensen}\ and\ \citenamefont
  {Cahay}(2006)}]{Jensen2006}%
  \BibitemOpen
  \bibfield  {author} {\bibinfo {author} {\bibfnamefont {K.~L.}\ \bibnamefont
  {Jensen}}\ and\ \bibinfo {author} {\bibfnamefont {M.}~\bibnamefont {Cahay}},\
  }\href {\doibase 10.1063/1.2193776} {\bibfield  {journal} {\bibinfo
  {journal} {Applied Physics Letters}\ }\textbf {\bibinfo {volume} {88}},\
  \bibinfo {pages} {154105} (\bibinfo {year} {2006})}\BibitemShut {NoStop}%
\bibitem [{\citenamefont {He}\ \emph {et~al.}(1991)\citenamefont {He},
  \citenamefont {Cutler},\ and\ \citenamefont {Miskovsky}}]{CutlerAPL}%
  \BibitemOpen
  \bibfield  {author} {\bibinfo {author} {\bibfnamefont {J.}~\bibnamefont
  {He}}, \bibinfo {author} {\bibfnamefont {P.}~\bibnamefont {Cutler}}, \ and\
  \bibinfo {author} {\bibfnamefont {N.}~\bibnamefont {Miskovsky}},\ }\href
  {\doibase 10.1063/1.106257} {\bibfield  {journal} {\bibinfo  {journal}
  {Applied Physics Letters}\ }\textbf {\bibinfo {volume} {59}},\ \bibinfo
  {pages} {1644} (\bibinfo {year} {1991})}\BibitemShut {NoStop}%
\bibitem [{\citenamefont {Kyritsakis}\ and\ \citenamefont
  {Xanthakis}(2015)}]{KXnonfn}%
  \BibitemOpen
  \bibfield  {author} {\bibinfo {author} {\bibfnamefont {A.}~\bibnamefont
  {Kyritsakis}}\ and\ \bibinfo {author} {\bibfnamefont {J.~P.}\ \bibnamefont
  {Xanthakis}},\ }\href {\doibase 10.1098/rspa.2014.0811} {\bibfield  {journal}
  {\bibinfo  {journal} {Proceedings of the Royal Society A}\ }\textbf {\bibinfo
  {volume} {471}},\ \bibinfo {pages} {20140811} (\bibinfo {year}
  {2015})}\BibitemShut {NoStop}%
\bibitem [{\citenamefont {Kyritsakis}\ and\ \citenamefont
  {Xanthakis}(2016)}]{KXGTF}%
  \BibitemOpen
  \bibfield  {author} {\bibinfo {author} {\bibfnamefont {A.}~\bibnamefont
  {Kyritsakis}}\ and\ \bibinfo {author} {\bibfnamefont {J.}~\bibnamefont
  {Xanthakis}},\ }\href {\doibase 10.1063/1.4940721} {\bibfield  {journal}
  {\bibinfo  {journal} {Journal of Applied Physics}\ }\textbf {\bibinfo
  {volume} {119}} (\bibinfo {year} {2016}),\ 10.1063/1.4940721}\BibitemShut
  {NoStop}%
\bibitem [{\citenamefont {Kyritsakis}\ and\ \citenamefont
  {Djurabekova}(2017)}]{GETELECpaper}%
  \BibitemOpen
  \bibfield  {author} {\bibinfo {author} {\bibfnamefont {A.}~\bibnamefont
  {Kyritsakis}}\ and\ \bibinfo {author} {\bibfnamefont {F.}~\bibnamefont
  {Djurabekova}},\ }\href {\doibase 10.1016/j.commatsci.2016.11.010} {\bibfield
   {journal} {\bibinfo  {journal} {Computational Materials Science}\ }\textbf
  {\bibinfo {volume} {128}},\ \bibinfo {pages} {15} (\bibinfo {year}
  {2017})}\BibitemShut {NoStop}%
\bibitem [{\citenamefont {Child}(1911)}]{Child_SC}%
  \BibitemOpen
  \bibfield  {author} {\bibinfo {author} {\bibfnamefont {C.~D.}\ \bibnamefont
  {Child}},\ }\href {\doibase 10.1103/PhysRevSeriesI.32.492} {\bibfield
  {journal} {\bibinfo  {journal} {Physical Review (Series I)}\ }\textbf
  {\bibinfo {volume} {32}},\ \bibinfo {pages} {492} (\bibinfo {year}
  {1911})}\BibitemShut {NoStop}%
\bibitem [{\citenamefont {Jensen}(1999)}]{Jensen1999}%
  \BibitemOpen
  \bibfield  {author} {\bibinfo {author} {\bibfnamefont {K.~L.}\ \bibnamefont
  {Jensen}},\ }in\ \href {\doibase 10.1002/047134608X.W3129.pub2} {\emph
  {\bibinfo {booktitle} {Wiley Encyclopedia of Electrical and Electronics
  Engineering}}}\ (\bibinfo  {publisher} {John Wiley \& Sons Inc.},\ \bibinfo
  {year} {1999})\BibitemShut {NoStop}%
\bibitem [{\citenamefont {Chen}\ \emph {et~al.}(2009)\citenamefont {Chen},
  \citenamefont {Cheng}, \citenamefont {Tsai},\ and\ \citenamefont
  {Shao}}]{chen2009space}%
  \BibitemOpen
  \bibfield  {author} {\bibinfo {author} {\bibfnamefont {P.}~\bibnamefont
  {Chen}}, \bibinfo {author} {\bibfnamefont {T.}~\bibnamefont {Cheng}},
  \bibinfo {author} {\bibfnamefont {J.}~\bibnamefont {Tsai}}, \ and\ \bibinfo
  {author} {\bibfnamefont {Y.}~\bibnamefont {Shao}},\ }\href@noop {} {\bibfield
   {journal} {\bibinfo  {journal} {Nanotechnology}\ }\textbf {\bibinfo {volume}
  {20}},\ \bibinfo {pages} {405202} (\bibinfo {year} {2009})}\BibitemShut
  {NoStop}%
\bibitem [{\citenamefont {Uimanov}(2011)}]{Uimanov2011}%
  \BibitemOpen
  \bibfield  {author} {\bibinfo {author} {\bibfnamefont {I.~V.}\ \bibnamefont
  {Uimanov}},\ }\href {\doibase 10.1109/TDEI.2011.5931082} {\bibfield
  {journal} {\bibinfo  {journal} {IEEE Transactions on Dielectrics and
  Electrical Insulation}\ }\textbf {\bibinfo {volume} {18}},\ \bibinfo {pages}
  {924} (\bibinfo {year} {2011})}\BibitemShut {NoStop}%
\bibitem [{\citenamefont {Forbes}(2008)}]{ForbesSpace}%
  \BibitemOpen
  \bibfield  {author} {\bibinfo {author} {\bibfnamefont {R.~G.}\ \bibnamefont
  {Forbes}},\ }\href {\doibase 10.1063/1.2996005} {\bibfield  {journal}
  {\bibinfo  {journal} {Journal of Applied Physics}\ }\textbf {\bibinfo
  {volume} {104}},\ \bibinfo {pages} {084303} (\bibinfo {year}
  {2008})}\BibitemShut {NoStop}%
\bibitem [{\citenamefont {Barbour}\ \emph {et~al.}(1953)\citenamefont
  {Barbour}, \citenamefont {Dolan}, \citenamefont {Trolan}, \citenamefont
  {Martin},\ and\ \citenamefont {Dyke}}]{BarbourSC}%
  \BibitemOpen
  \bibfield  {author} {\bibinfo {author} {\bibfnamefont {J.}~\bibnamefont
  {Barbour}}, \bibinfo {author} {\bibfnamefont {W.}~\bibnamefont {Dolan}},
  \bibinfo {author} {\bibfnamefont {J.}~\bibnamefont {Trolan}}, \bibinfo
  {author} {\bibfnamefont {E.}~\bibnamefont {Martin}}, \ and\ \bibinfo {author}
  {\bibfnamefont {W.}~\bibnamefont {Dyke}},\ }\href {\doibase
  10.1103/PhysRev.92.45} {\bibfield  {journal} {\bibinfo  {journal} {Physical
  Review}\ }\textbf {\bibinfo {volume} {92}},\ \bibinfo {pages} {45} (\bibinfo
  {year} {1953})}\BibitemShut {NoStop}%
\bibitem [{\citenamefont {Bejan}\ and\ \citenamefont
  {Kraus}(2003)}]{bejan2003heat}%
  \BibitemOpen
  \bibfield  {author} {\bibinfo {author} {\bibfnamefont {A.}~\bibnamefont
  {Bejan}}\ and\ \bibinfo {author} {\bibfnamefont {A.~D.}\ \bibnamefont
  {Kraus}},\ }\href@noop {} {\emph {\bibinfo {title} {Heat transfer
  handbook}}},\ Vol.~\bibinfo {volume} {1}\ (\bibinfo  {publisher} {John Wiley
  \& Sons},\ \bibinfo {year} {2003})\BibitemShut {NoStop}%
\bibitem [{\citenamefont {Matula}(1979)}]{Matula1979}%
  \BibitemOpen
  \bibfield  {author} {\bibinfo {author} {\bibfnamefont {R.~A.}\ \bibnamefont
  {Matula}},\ }\href {\doibase 10.1063/1.555614} {\bibfield  {journal}
  {\bibinfo  {journal} {Journal of Physical and Chemical Reference Data}\
  }\textbf {\bibinfo {volume} {8}},\ \bibinfo {pages} {1147} (\bibinfo {year}
  {1979})}\BibitemShut {NoStop}%
\bibitem [{\citenamefont {Gathers}(1983)}]{Gathers1983}%
  \BibitemOpen
  \bibfield  {author} {\bibinfo {author} {\bibfnamefont {G.~R.}\ \bibnamefont
  {Gathers}},\ }\href {\doibase 10.1007/BF00502353} {\bibfield  {journal}
  {\bibinfo  {journal} {International Journal of Thermophysics}\ }\textbf
  {\bibinfo {volume} {4}},\ \bibinfo {pages} {209} (\bibinfo {year}
  {1983})}\BibitemShut {NoStop}%
\bibitem [{\citenamefont {Yarimbiyik}\ \emph {et~al.}(2006)\citenamefont
  {Yarimbiyik}, \citenamefont {Schafft}, \citenamefont {Allen}, \citenamefont
  {Zaghloul},\ and\ \citenamefont {Blackburn}}]{Yarimbiyik2005}%
  \BibitemOpen
  \bibfield  {author} {\bibinfo {author} {\bibfnamefont {A.~E.}\ \bibnamefont
  {Yarimbiyik}}, \bibinfo {author} {\bibfnamefont {H.~A.}\ \bibnamefont
  {Schafft}}, \bibinfo {author} {\bibfnamefont {R.~A.}\ \bibnamefont {Allen}},
  \bibinfo {author} {\bibfnamefont {M.~E.}\ \bibnamefont {Zaghloul}}, \ and\
  \bibinfo {author} {\bibfnamefont {D.~L.}\ \bibnamefont {Blackburn}},\ }\href
  {\doibase https://doi.org/10.1016/j.microrel.2005.09.004} {\bibfield
  {journal} {\bibinfo  {journal} {Microelectronics Reliability}\ }\textbf
  {\bibinfo {volume} {46}},\ \bibinfo {pages} {1050 } (\bibinfo {year}
  {2006})}\BibitemShut {NoStop}%
\bibitem [{\citenamefont {Nath}\ and\ \citenamefont {Chopra}(1974)}]{Nath1974}%
  \BibitemOpen
  \bibfield  {author} {\bibinfo {author} {\bibfnamefont {P.}~\bibnamefont
  {Nath}}\ and\ \bibinfo {author} {\bibfnamefont {K.}~\bibnamefont {Chopra}},\
  }\href {\doibase 10.1016/0040-6090(74)90033-9} {\bibfield  {journal}
  {\bibinfo  {journal} {Thin Solid Films}\ }\textbf {\bibinfo {volume} {20}},\
  \bibinfo {pages} {53} (\bibinfo {year} {1974})}\BibitemShut {NoStop}%
\bibitem [{\citenamefont {Berendsen}\ \emph {et~al.}(1984)\citenamefont
  {Berendsen}, \citenamefont {Postma}, \citenamefont {van Gunsteren},
  \citenamefont {DiNola},\ and\ \citenamefont {Haak}}]{Berendsen}%
  \BibitemOpen
  \bibfield  {author} {\bibinfo {author} {\bibfnamefont {H.~J.~C.}\
  \bibnamefont {Berendsen}}, \bibinfo {author} {\bibfnamefont {J.~P.~M.}\
  \bibnamefont {Postma}}, \bibinfo {author} {\bibfnamefont {W.~F.}\
  \bibnamefont {van Gunsteren}}, \bibinfo {author} {\bibfnamefont
  {A.}~\bibnamefont {DiNola}}, \ and\ \bibinfo {author} {\bibfnamefont {J.~R.}\
  \bibnamefont {Haak}},\ }\href {\doibase http://dx.doi.org/10.1063/1.448118}
  {\bibfield  {journal} {\bibinfo  {journal} {The Journal of Chemical Physics}\
  }\textbf {\bibinfo {volume} {81}},\ \bibinfo {pages} {3684} (\bibinfo {year}
  {1984})}\BibitemShut {NoStop}%
\bibitem [{\citenamefont {Wu}\ \emph {et~al.}(2017)\citenamefont {Wu},
  \citenamefont {Shi}, \citenamefont {Chen}, \citenamefont {Shao},
  \citenamefont {Abe}, \citenamefont {Higo}, \citenamefont {Matsumoto},\ and\
  \citenamefont {Wuensch}}]{Wu_high-gradient}%
  \BibitemOpen
  \bibfield  {author} {\bibinfo {author} {\bibfnamefont {X.}~\bibnamefont
  {Wu}}, \bibinfo {author} {\bibfnamefont {J.}~\bibnamefont {Shi}}, \bibinfo
  {author} {\bibfnamefont {H.}~\bibnamefont {Chen}}, \bibinfo {author}
  {\bibfnamefont {J.}~\bibnamefont {Shao}}, \bibinfo {author} {\bibfnamefont
  {T.}~\bibnamefont {Abe}}, \bibinfo {author} {\bibfnamefont {T.}~\bibnamefont
  {Higo}}, \bibinfo {author} {\bibfnamefont {S.}~\bibnamefont {Matsumoto}}, \
  and\ \bibinfo {author} {\bibfnamefont {W.}~\bibnamefont {Wuensch}},\ }\href
  {\doibase 10.1103/PhysRevAccelBeams.20.052001} {\bibfield  {journal}
  {\bibinfo  {journal} {Physical Review Special Topics - Accelerators and
  Beams}\ }\textbf {\bibinfo {volume} {20}},\ \bibinfo {pages} {052001}
  (\bibinfo {year} {2017})}\BibitemShut {NoStop}%
\bibitem [{\citenamefont {Sabochick}\ and\ \citenamefont
  {Lam}(1991)}]{SL_EAM_CU}%
  \BibitemOpen
  \bibfield  {author} {\bibinfo {author} {\bibfnamefont {M.~J.}\ \bibnamefont
  {Sabochick}}\ and\ \bibinfo {author} {\bibfnamefont {N.~Q.}\ \bibnamefont
  {Lam}},\ }\href {\doibase 10.1103/PhysRevB.43.5243} {\bibfield  {journal}
  {\bibinfo  {journal} {Physical Review B}\ }\textbf {\bibinfo {volume} {43}},\
  \bibinfo {pages} {5243} (\bibinfo {year} {1991})}\BibitemShut {NoStop}%
\bibitem [{\citenamefont {Mishin}\ \emph {et~al.}(2001)\citenamefont {Mishin},
  \citenamefont {Mehl}, \citenamefont {Papaconstantopoulos}, \citenamefont
  {Voter},\ and\ \citenamefont {Kress}}]{Mishin}%
  \BibitemOpen
  \bibfield  {author} {\bibinfo {author} {\bibfnamefont {Y.}~\bibnamefont
  {Mishin}}, \bibinfo {author} {\bibfnamefont {M.~J.}\ \bibnamefont {Mehl}},
  \bibinfo {author} {\bibfnamefont {D.~A.}\ \bibnamefont
  {Papaconstantopoulos}}, \bibinfo {author} {\bibfnamefont {A.~F.}\
  \bibnamefont {Voter}}, \ and\ \bibinfo {author} {\bibfnamefont {J.~D.}\
  \bibnamefont {Kress}},\ }\href {\doibase 10.1103/PhysRevB.63.224106}
  {\bibfield  {journal} {\bibinfo  {journal} {Physical Review B}\ }\textbf
  {\bibinfo {volume} {63}},\ \bibinfo {pages} {224106} (\bibinfo {year}
  {2001})}\BibitemShut {NoStop}%
\bibitem [{\citenamefont {Ester}\ \emph {et~al.}(1996)\citenamefont {Ester},
  \citenamefont {Kriegel}, \citenamefont {Sander}, \citenamefont {Xu} \emph
  {et~al.}}]{ester1996density}%
  \BibitemOpen
  \bibfield  {author} {\bibinfo {author} {\bibfnamefont {M.}~\bibnamefont
  {Ester}}, \bibinfo {author} {\bibfnamefont {H.-P.}\ \bibnamefont {Kriegel}},
  \bibinfo {author} {\bibfnamefont {J.}~\bibnamefont {Sander}}, \bibinfo
  {author} {\bibfnamefont {X.}~\bibnamefont {Xu}},  \emph {et~al.},\ }in\
  \href@noop {} {\emph {\bibinfo {booktitle} {Kdd}}},\ Vol.~\bibinfo {volume}
  {96}\ (\bibinfo {year} {1996})\ pp.\ \bibinfo {pages} {226--231}\BibitemShut
  {NoStop}%
\bibitem [{\citenamefont {Batrakov}\ \emph {et~al.}(1999)\citenamefont
  {Batrakov}, \citenamefont {Proskurovsky},\ and\ \citenamefont
  {Popov}}]{Batrakov_melt}%
  \BibitemOpen
  \bibfield  {author} {\bibinfo {author} {\bibfnamefont {A.}~\bibnamefont
  {Batrakov}}, \bibinfo {author} {\bibfnamefont {D.}~\bibnamefont
  {Proskurovsky}}, \ and\ \bibinfo {author} {\bibfnamefont {S.}~\bibnamefont
  {Popov}},\ }\href {\doibase 10.1109/94.788735} {\bibfield  {journal}
  {\bibinfo  {journal} {IEEE Transactions on Dielectrics and Electrical
  Insulation}\ }\textbf {\bibinfo {volume} {6}},\ \bibinfo {pages} {410}
  (\bibinfo {year} {1999})}\BibitemShut {NoStop}%
\bibitem [{\citenamefont {Kildemo}(2004)}]{CERN2004}%
  \BibitemOpen
  \bibfield  {author} {\bibinfo {author} {\bibfnamefont {M.}~\bibnamefont
  {Kildemo}},\ }\href {\doibase 10.1016/j.nima.2004.04.230} {\bibfield
  {journal} {\bibinfo  {journal} {Nuclear Instruments and Methods in Physics
  Research A}\ }\textbf {\bibinfo {volume} {530}},\ \bibinfo {pages} {596}
  (\bibinfo {year} {2004})}\BibitemShut {NoStop}%
\bibitem [{\citenamefont {Veske}\ \emph
  {et~al.}(2016{\natexlab{b}})\citenamefont {Veske}, \citenamefont
  {Parviainen}, \citenamefont {Zadin}, \citenamefont {Aabloo},\ and\
  \citenamefont {Djurabekova}}]{Veske2016}%
  \BibitemOpen
  \bibfield  {author} {\bibinfo {author} {\bibfnamefont {M.}~\bibnamefont
  {Veske}}, \bibinfo {author} {\bibfnamefont {S.}~\bibnamefont {Parviainen}},
  \bibinfo {author} {\bibfnamefont {V.}~\bibnamefont {Zadin}}, \bibinfo
  {author} {\bibfnamefont {A.}~\bibnamefont {Aabloo}}, \ and\ \bibinfo {author}
  {\bibfnamefont {F.}~\bibnamefont {Djurabekova}},\ }\href {\doibase
  10.1088/0022-3727/49/21/215301} {\bibfield  {journal} {\bibinfo  {journal}
  {Journal of Physics D: Applied Physics}\ }\textbf {\bibinfo {volume} {49}},\
  \bibinfo {pages} {215301} (\bibinfo {year} {2016}{\natexlab{b}})}\BibitemShut
  {NoStop}%
\bibitem [{\citenamefont {Jansson}\ \emph {et~al.}(2016)\citenamefont
  {Jansson}, \citenamefont {Baibuz},\ and\ \citenamefont
  {Djurabekova}}]{janssonKMC}%
  \BibitemOpen
  \bibfield  {author} {\bibinfo {author} {\bibfnamefont {V.}~\bibnamefont
  {Jansson}}, \bibinfo {author} {\bibfnamefont {E.}~\bibnamefont {Baibuz}}, \
  and\ \bibinfo {author} {\bibfnamefont {F.}~\bibnamefont {Djurabekova}},\
  }\href@noop {} {\bibfield  {journal} {\bibinfo  {journal} {Nanotechnology}\
  }\textbf {\bibinfo {volume} {27}},\ \bibinfo {pages} {265708} (\bibinfo
  {year} {2016})}\BibitemShut {NoStop}%
\end{thebibliography}%

\end{document}